\documentclass[10pt, journal, compsoc, twoside, hidelinks]{IEEEtran}
\usepackage[utf8]{inputenc}
\usepackage[T1]{fontenc}
\usepackage[english]{babel}

\usepackage[nocompress]{cite}
\newcommand{\ignore}[1]{}
\usepackage{fancyhdr}
\usepackage[normalem]{ulem}
\usepackage[hyphens]{url}
\usepackage{booktabs}
\usepackage{xfrac}
\usepackage{algorithm}
\usepackage{algorithmic}
\usepackage{xspace}
\usepackage{tabularx}
\usepackage{placeins}
\usepackage[para]{threeparttable}
\usepackage{multirow}
\usepackage{multicol}
\usepackage{color}
\usepackage{tikz}
\usepackage{amssymb}
\usepackage{pifont}
\usepackage[inline]{enumitem}
\usepackage{colortbl}
\usepackage{amsmath}
\usepackage[draft]{hyperref}
\usepackage{listings}
\usepackage{etoolbox}
\usepackage[printwatermark]{xwatermark}
\usepackage{mwe}
\usepackage[shortcuts]{extdash}
\usepackage[acronym]{glossaries}
\usepackage{xspace}
\usepackage{tabularx}
\usepackage{booktabs}
\usepackage{multirow}
\usepackage{tikz}

\usepackage[binary-units=true, per-mode=symbol]{siunitx}
\DeclareSIUnit\FLOP{flop}
\DeclareSIUnit\DFLOP{DPflop}
\DeclareSIUnit\cycle{cycle}
\DeclareSIUnit\FLOPs{\FLOP\per\second}
\DeclareSIUnit\FLOPsW{\FLOPs\per\watt}
\DeclareSIUnit\GFLOPs{\giga\FLOPs}
\DeclareSIUnit\DGFLOPs{DP\giga\FLOPs}
\DeclareSIUnit\SGFLOPs{SP\giga\FLOPs}
\DeclareSIUnit\GFLOPsW{\giga\FLOPsW}
\DeclareSIUnit\DGFLOPsW{DP\giga\FLOPsW}
\DeclareSIUnit\SGFLOPsW{SP\giga\FLOPsW}
\DeclareSIUnit\TFLOPs{\tera\FLOPs}
\DeclareSIUnit\TFLOPsW{\tera\FLOPsW}
\DeclareSIUnit\DTFLOPsW{DP\tera\FLOPsW}
\DeclareSIUnit\STFLOPsW{SP\tera\FLOPsW}
\DeclareSIUnit\EFLOPs{\exa\FLOPs}
\DeclareSIUnit\gateequivalent {GE}
\DeclareSIUnit\kGE {\kilo\gateequivalent{}}
\DeclareSIUnit\MGE {\mega\gateequivalent{}}
\DeclareSIUnit\cycle {cycle}

\newcolumntype{C}{>{\centering\arraybackslash}X}
\newcolumntype{R}{>{\raggedleft\arraybackslash}X}
\newcolumntype{x}[1]{>{\centering\arraybackslash\hspace{0pt}}p{#1}}
\newcolumntype{k}[1]{S[
      table-format=#1,
      detect-weight,
      input-symbols={()},
    ]}

\def\reviewpass{7}

\newcommand{\secref}[1]{{Section~\ref{#1}}}

\newcommand{\figref}[1]{{Figure~\ref{#1}}}
\newcommand{\tabref}[1]{{Table~\ref{#1}}}

\newacronym{asic}{ASIC}{Application-Specific Integrated Circuit}
\newacronym{tpfpu}{TP\=/FPU}{transprecision floating-point unit}
\newacronym{fpu}{FPU}{floating\=/point unit}
\newacronym{fp}{FP}{floating\=/point}
\newacronym{simd}{SIMD}{single instruction multiple data}
\newacronym{simt}{SIMT}{single instruction multiple thread}
\newacronym{hpc}{HPC}{High Performance Computing}
\newacronym{iot}{IoT}{Internet of Things}
\newacronym{fma}{FMA}{fused multiply-add}
\newacronym{tpu}{TPU}{Tensor Processing Unit}
\newacronym{ml}{ML}{machine learning}
\newacronym{tdp}{TDP}{Thermal Design Power}
\newacronym{bb}{BB}{body bias}
\newacronym{fbb}{FBB}{forward body bias}
\newacronym{ntx}{NTX}{network training accelerator}
\newacronym{axi}{AXI}{Advanced eXtensible Interface}
\newacronym{uart}{UART}{Universal Asynchronous Receiver Transmitter}
\newacronym{ahp}{AHP}{Ariane High Performance}
\newacronym{alp}{ALP}{Ariane Low Power}
\newacronym{fll}{FLL}{frequency locked loop}
\newacronym{alu}{ALU}{arithmetic logic unit}
\newacronym{lsu}{LSU}{load store unit}
\newacronym[longplural={static random-access memories}]{sram}{SRAM}{static random-access memory}
\newacronym{os}{OS}{operating system}
\newacronym{dibl}{DIBL}{drain-induced barrier lowering}
\newacronym{soc}{SoC}{system on chip}
\newacronym{cpu}{CPU}{central processing unit}
\newacronym{gpu}{GPU}{graphics processing unit}
\newacronym{isa}{ISA}{instruction set architecture}
\newacronym{cc}{CC}{core complex}
\newacronym{csr}{CSR}{control and status register}
\newacronym{tcdm}{TCDM}{tightly coupled data memory}
\newacronym{spm}{SPM}{scratchpad memory}
\newacronym{fpss}{FP-SS}{floating point subsystem}
\newacronym{ssr}{SSR}{stream semantic register}
\newacronym{frep}{FREP}{floating-point repetition instruction}
\newacronym{lrsc}{LR/SC}{load-reserved/store-conditional}
\newacronym{fsm}{FSM}{finite-state machine}
\newacronym{ipi}{IPI}{inter-processor interrupt}
\newacronym{io}{IO}{input/output}
\newacronym{sta}{STA}{static timing analysis}
\newacronym{fft}{FFT}{Fast Fourier Transform}
\newacronym{relu}{ReLU}{rectified linear unit}
\newacronym{rtl}{RTL}{register transfer level}
\newacronym{ipc}{IPC}{instructions per cycle}
\newacronym{cpi}{CPI}{cycles per instruction}
\newacronym{rf}{RF}{register file}
\newacronym{dma}{DMA}{direct memory access}
\newacronym{mosfet}{MOSFET}{metal–oxide–semiconductor field-effect transistor}
\newacronym{pmc}{PMC}{performance monitoring counter}
\newacronym{ptw}{PTW}{page table walker}

\newacronym{sse}{SSE}{Streaming SIMD extensions}
\newacronym{avx}{AVX}{advanced vector extensions}
\newacronym{neon}{NEON}{NEON Media Processing Engine}
\newacronym{sve}{SVE}{scaleable vector extensions}
\newacronym{arm}{Arm}{Advanced RISC Machines}
\newacronym{gpgpu}{GPGPU}{General Purpose Computation on Graphics Processing Unit}
\newacronym{sm}{SM}{streaming multiprocessor}
\newacronym{ilp}{ILP}{instruction level parallelism}
\newacronym{vl}{VL}{vector length}
\newacronym{llvm}{LLVM}{Low Level Virtual Machine}

\newacronym{vrf}{VRF}{vector register file}
\newacronym{vpu}{VPU}{vector processing unit}
\newacronym{dsp}{DSP}{digital signal processor}
\newacronym{sfu}{SFU}{special function unit}
\newacronym{blas}{blas}{basic linear algebra subprograms}

\newcommand{\fpstd}{IEEE\=/754\xspace}
\newcommand{\fp}{floating\=/point\xspace}
\newcommand{\riscv}{RISC\=/V\xspace}
\newcommand{\arm}{Arm\xspace}
\newcommand{\gf}{\textsc{Glo\-bal\-found\-ries}~22\,nm~FDX\xspace}

\newcommand{\designcompiler}{\textsc{Synopsys~Design~Compiler~2017.09}\xspace}
\newcommand{\primetime}{\textsc{Synopsys~Primetime~2019.12}\xspace}
\newcommand{\innovus}{\textsc{Cadence~Innovus~17.11}\xspace}

\newcommand{\vtr}{valid-then-ready\xspace}
\newcommand{\frep}{\texttt{frep}\xspace}

\newcommand{\todo}[2]{%
  \ifnum\reviewpass>#1{%
    \unskip\ignorespaces%
    }\else{%
    {\color{red}{\bf [#2]}}%
    }\fi%
}

\definecolor{revision}{rgb}{0.0, 0.4, 0.8}
\def\revcolor{revision}

\graphicspath{{fig/}}

\lstdefinelanguage{rvasm}{
  sensitive = true, 
  morecomment = [l]{;}, 
}

\begin{document}
%
\title{Snitch: A tiny Pseudo Dual-Issue Processor for Area and Energy Efficient Execution of Floating-Point Intensive Workloads}
%
%
%
%

\author{Florian~Zaruba,
  Fabian~Schuiki,
  Torsten~Hoefler,
  and~Luca~Benini
  \IEEEcompsocitemizethanks{\IEEEcompsocthanksitem F. Zaruba, F. Schuiki and L. Benini are with the Integrated Systems Laboratory (IIS), Swiss Federal Institute of Technology, Zurich, Switzerland\protect\\
    E-mail: \{zarubaf,fschuiki,benini\}@iis.ee.ethz.ch
    \IEEEcompsocthanksitem T. Hoefler is with the Scalable Parallel Computing Laboratory (SPCL), Swiss Federal Institute of Technology, Zurich, Switzerland\protect\\
    E-mail: htor@inf.ethz.ch
    \IEEEcompsocthanksitem L. Benini also is with Department of Electrical, Electronic and Information Engineering (DEI), University of Bologna, Bologna, Italy.}
}

%
%

\markboth{IEEE TRANSACTIONS ON COMPUTERS, VOL. (VOL), NO. (NO), (MONTH) (YEAR)}%
{Zaruba \MakeLowercase{\textit{et al.}}: Area and Energy efficient architecture for floating-point workloads}
%



\IEEEtitleabstractindextext{%
  \begin{abstract}
    Data-parallel applications, such as data analytics, machine learning, and scientific computing, are placing an ever-growing demand on floating-point operations per second on emerging systems. With increasing integration density, the quest for energy efficiency becomes the number one design concern. While dedicated accelerators provide high energy efficiency, they are over-specialized and hard to adjust to algorithmic changes. We propose an architectural concept that tackles the issues of achieving extreme energy efficiency while still maintaining high flexibility as a general-purpose compute engine. The key idea is to pair a tiny \SI{10}{\kGE} (kilo gate equivalent) control core, called Snitch, with a double-precision \acrfull{fpu} to adjust the compute to control ratio. While traditionally minimizing non-\gls{fpu} area and achieving high \fp utilization has been a trade-off, with Snitch, we achieve them both, by enhancing the ISA with two minimally intrusive extensions: stream semantic registers (SSR) and a floating-point repetition instruction (FREP). SSRs allow the core to implicitly encode load/store instructions as register reads/writes, eliding many explicit memory instructions. The FREP extension decouples the floating-point and integer pipeline by sequencing instructions from a micro-loop buffer. These ISA extensions significantly reduce the pressure on the core and free it up for other tasks, making Snitch and FPU effectively dual-issue at a minimal incremental cost of 3.2\%. The two low overhead ISA extensions make Snitch more flexible than a contemporary vector processor lane, achieving a $2\times$ energy-efficiency improvement. We have evaluated the proposed core and ISA extensions on an octa-core cluster in 22\,nm technology. We achieve more than $6\times$ multi-core speed-up and a $3.5\times$ gain in energy efficiency on several parallel microkernels.
  \end{abstract}

  }

\IEEEoverridecommandlockouts
\IEEEpubid{\makebox[\columnwidth]{0018-9340~\copyright2020 IEEE \hfill} \hspace{\columnsep}\makebox[\columnwidth]{ }}
\maketitle
\IEEEpubidadjcol

\IEEEdisplaynontitleabstractindextext

%
\IEEEpeerreviewmaketitle

\IEEEraisesectionheading{\section{Introduction}\label{sec:introduction}}

\IEEEPARstart{T}{he} ever-increasing demand for floating-point performance in scientific computing, machine learning, big data analytics, and human-computer interaction are dominating the requirements for next-generation computer systems~\cite{yao2019pursuing}. The paramount design goal to satisfy the demand of computing resources is energy efficiency: Shrinking feature sizes allow us to pack billions of transistors in dies as large as \SI{600}{\square\milli\meter}. The high transistor density makes it impossible to switch all of them at the same time at high speed as the consumed power in the form of heat cannot dissipate into the environment fast enough. Ultimately, designers have to be more careful than ever only to spend energy on logic, which contributes to solving the problem.

Thus, we see an explosion on the number of accelerators solely dedicated to solving one particular problem efficiently. Unfortunately, there is only a limited optimization space that, with the end of technology scaling, will reach a limit of a near-optimal hardware architecture for a certain problem~\cite{fuchs2019accelerator}. Furthermore, algorithms can evolve rapidly, thereby making domain-specific architectures less efficient for such algorithms~\cite{nowatzki2016pushing}.
On the other end of the spectrum, we can find fully programmable systems such as \glspl{gpu} and even more general-purpose units like \glspl{cpu}. The programmability and flexibility of those systems incur significant overhead and make such systems less energy efficient. Furthermore, \glspl{cpu} and \glspl{gpu} (to a lesser degree) are affected by the \emph{Von Neumann bottleneck}: The rate of which information can travel from data and instruction memory limits the computational throughput of the architecture. More hardware is necessary to mitigate these effects, such as caching, multi-threading, and super-scalar out-of-order processor pipelines~\cite{hennessy2011computer}. All these mitigation techniques aim to increase the \emph{utilization} of the compute resource, in this case, the \gls{fpu}. They achieve this goal at a price of much-increased hardware complexity, which in turn decreases efficiency because a smaller part of the silicon budget remains dedicated to compute units. A reproducible example, thanks due to its open-source nature, is the out-of-order BOOM \gls{cpu}~\cite{celio2017boom,zhaosonicboom}: Approximately \SI{2.7}{\percent} of the core's overall area, is spent on the \gls{fpu}.\footnote{estimated on a post-synthesis netlist in \gf} More advanced \glspl{cpu} such as AMD's Zen2 architecture show a better compute per area efficiency (around \SI{25}{\percent}), primarily thanks to the wide \gls{simd} \fp execution units~\cite{zen22020isscc}.

\subsection{Design Goal: Area and Energy efficiency}

\begin{figure}
    \centering
    \includegraphics[width=\linewidth]{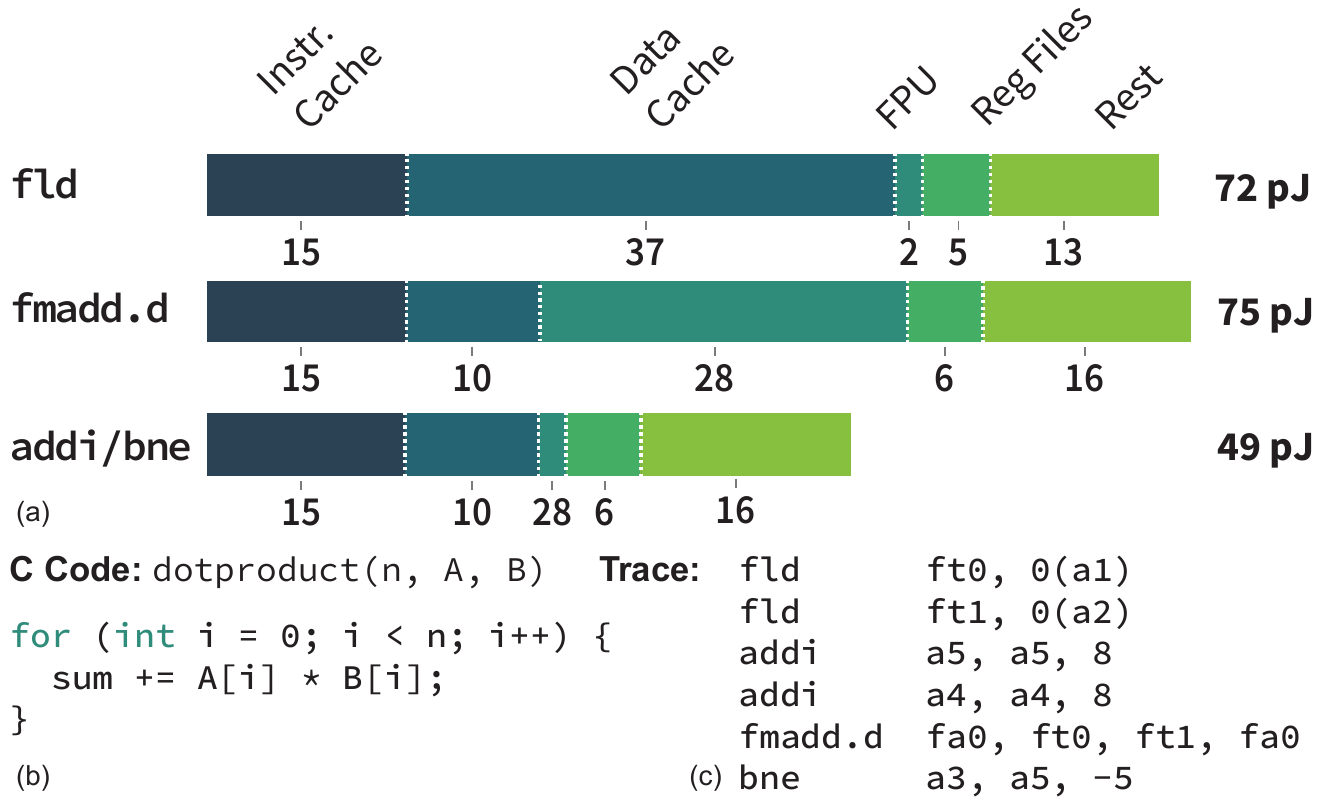}
    \caption{(a) Energy per instruction [\si{\pico\joule}] for instructions used in a simple dot product kernel. (b) corresponding C code and (c) (simplified) \riscv assembly. Two load instructions, one floating-point accumulate instruction, and one branch instruction make up the inner-most loop kernel. We provide energy per op of an application-class \riscv processor called Ariane~\cite{zaruba2019ariane}. In total one loop operation consumes \SI{317}{\pico\joule} for which only \SI{28}{\pico\joule} are spent on the actual computation.}
    \label{fig:energy_per_instruction}
\end{figure}
To give the reader quantitative intuition on the severe efficiency limits affecting programmable architectures, let us consider, the simple kernel of a dot product ($z = \vec{a} \cdot \vec{b}$) in \figref{fig:energy_per_instruction}(b,c). The corresponding energies per instruction type in \figref{fig:energy_per_instruction}(a) for a 64-bit application-class \riscv processor as reported in~\cite{zaruba2019ariane} in a \SI{22}{\nano\meter} technology.
The kernel consists of up of five instructions. Four of those instructions perform bookkeeping tasks such as moving the data into the local \gls{rf} on which arithmetic instructions can operate and looping over all $n$ elements of the input vectors. In total, the energy used for performing an element multiplication and addition in this setting is \SI{317}{\pico\joule}. The only ``useful'' workload in this kernel is performed by the \gls{fpu}, which accounts for \SI{28}{\pico\joule}. The rest of the energy (\SI{289}{\pico\joule}) is spent on auxiliary tasks. Even this short kernel gives us an immediate intuition on where energy efficiency is lost. \gls{fpu} utilization is low (\SI{17}{\percent}), mostly due to load and store and loop management instructions.
\subsection{Existing Mitigation Techniques and Architectures}
Techniques and architectures exist that try to mitigate the efficiency issue highlighted above.

\begin{itemize}
    \item \textbf{\Gls{isa} extensions}:
          Post increment load and store instruction can accelerate pointer bumping within a loop~\cite{gautschi2017near}. For an efficient implementation they require a second write-port into the \gls{rf}, therefore increasing the implementation cost. 
          \Gls{simd} such as \gls{sse}/\gls{avx}~\cite{cornea2015intel} in x86 or \gls{neon}~\cite{reddy2008neon} in \gls{arm} perform a single-instruction on a fixed amount of data items in a parallel fashion. Therefore reducing the total loop count and amortizing the loop overhead per computation. Wide \gls{simd} data-paths are quite inflexible when elements need to be accessed individually: Dedicated shuffle operations are used to bring the data into a \gls{simd}-amenable form. 
    \item \textbf{Vector architectures}:
          Cray-style~\cite{russell1978cray} vector units such as the \gls{sve}~\cite{stephens2017arm} and the \riscv vector extension~\cite{cavalcante2019ara} operate on larger chunks of data in the form of vectors.

    \item \textbf{\Glspl{gpu}}:
          \Gls{simt} architectures such as NVIDIA's V100~\cite{nvidia2017volta} \gls{gpu} use multiple parallel scalar threads that execute the same instructions. Hardware scheduling of threads hides memory latency. Coalescing units bundle the memory traffic to make accesses into (main) memory more efficient. However, the hardware to manage threads is quite complex and comes at a cost that offsets the energy efficiency of \glspl{gpu}. The thread scheduler needs to swap different thread contexts on the same \gls{sm} whenever it detects a stalling thread (group) waiting for memory loads to return or due to different outcomes of branches (branch divergence). This means that the \gls{sm} must keep a very large number of thread contexts (including the relatively large \gls{rf}) in local \glspl{sram}~\cite{jia2018dissecting}. \gls{sram} accesses incur a higher energy cost than reads to flipflop-based memories and enforce a word-wise access granularity. For \glspl{gpu} to overcome these limitations, they offer operand caches in which software can cache operands and results, which are then reusable at a later point in time, which decreases area and energy efficiency. For example, NVIDIA's Volta architecture offers two 64-bit read ports on its register file per thread. To sustain a three operand \gls{fma} instruction, it needs to source one operand from one of its operand caches~\cite{jia2018dissecting}. 
\end{itemize}

\subsection{Contributions}
\label{subsec:contributions}

The solutions we propose here to solve the problems outlined above are the following:
\begin{enumerate}
    \item A general-purpose, single-stage, single-issue core, called Snitch, tuned for high energy efficiency. Aiming to maximize the compute/control ratio (making the \gls{fpu} the dominant part of the design) mitigating the effects of deep pipelines and dynamic scheduling.
    \item An \gls{isa} extension, originally proposed by Schuiki \emph{et~al.}~\cite{schuiki2019stream}, called \gls{ssr}. This extension accelerates data-oblivious~\cite{goldreich1996software} problems by providing an efficient semantic to read and write from memory. Load and store instructions which follow affine access patterns (streams) are implicitly mapped to register read/writes. \Glspl{ssr} effectively elide all explicit memory operations. Semantically they are comparable to vector operations as they operate on vectors (tensors) without the explicit need for load and store instructions. We have enhanced the \gls{ssr} implementation by providing shadow registers to overlap configuration and computation. The shadow registers are transparent from a programming perspective, new configurations are accepted as long as the shadow registers are not full. As soon as the current configuration has finished, the shadow register's value is swapped in as a new active configuration. The streamers immediately start fetching using the new stream configuration.
    \item A second ISA extension, \gls{frep}, which controls an FPU sequence Buffer. The \gls{fpu} and the integer core in the proposed system are fully decoupled and only synchronize with explicit move instructions between the two subsystems. The \gls{fpu} sequencer is situated on the offloading path of the integer core to the \gls{fpu}. It provides a small, configurable size sequence buffer from which it can sequence \fp instructions in a configurable manner. The sequence buffer frees the integer core from issuing instructions to the \gls{fpu} that is, therefore, available for other control tasks. This makes this single-issue, in-order core \emph{pseudo dual-issue}, enabling it to overlap independent integer and \fp instructions. Furthermore, the sequence buffer eliminates the need for loops in the code and reduces the pressure on the instruction fetch. Repetition instructions are also implemented in the X86~\cite{intelisa2016} and TMS320C28x \gls{dsp}~\cite{tms320c28x2004cpu} \glspl{isa}. Compared to those instructions that allow only a single instruction to be repeated, our approach, in conjunction with \glspl{ssr}, offers greater flexibility as a few instruction can program the entire loop-buffer, and complex operations can be entirely offloaded.
\end{enumerate}
While traditionally minimizing non-\gls{fpu} area and achieving \fp high utilization has been a trade-off, we can eliminate the need to compromise: Our extensions have negligible area cost and boost FPU utilization significantly. Our Snitch core achieves the same clock frequency, higher flexibility, and is $2.0\times$ more area- and energy-efficient than a conventional vector processor lane.

From the design and implementation viewpoint, the contributions of this work are:
\begin{enumerate}
    \item A fully programmable, shared memory, multi-core system tuned for utmost energy efficiency by using a tiny integer core attached to a double-precision \gls{fpu}. Achieving $3.5\times$ more energy efficiency and $4.5\times$ better \gls{fpu} utilization on small matrices than the current state of the art.
    \item An implementation of the \gls{ssr}~\cite{schuiki2019stream} enhanced with shadow registers to allow overlapping loop-setup with ongoing operations using the \gls{frep} extension enabling the usage of our \gls{ssr} and \gls{frep} extensions on more irregular kernels such as \gls{fft}. Achieving speed-ups of $4.7\times$ in the single-core case and close to $3\times$ in the parallel octa-core case for the \gls{fft} benchmark.
    \item A decoupled \gls{fpu} and integer core architecture featuring a sequence buffer that can independently service the \gls{fpu} while the integer core is busy with control tasks. This extension, together with the \gls{ssr}, make the small integer core \emph{pseudo dual-issue} at a minimal incremental area cost of less than \SI{7}{\percent} for the core complex and \SI{3.2}{\percent} on the cluster level including memories.
\end{enumerate}

The rest of the paper is organized as follows: \secref{sec:architecture} describes the proposed architecture and \gls{isa} extensions, \secref{sec:programming_model} offers more details on the programming model of the system and the \gls{isa} extensions, \secref{sec:results} presents the experimental setup, evaluation and comparison to other systems. The last sections conclude the presented work and present future research directions.

\section{Architecture}
\label{sec:architecture}

\begin{figure*}
    \centering
    \includegraphics[width=\textwidth]{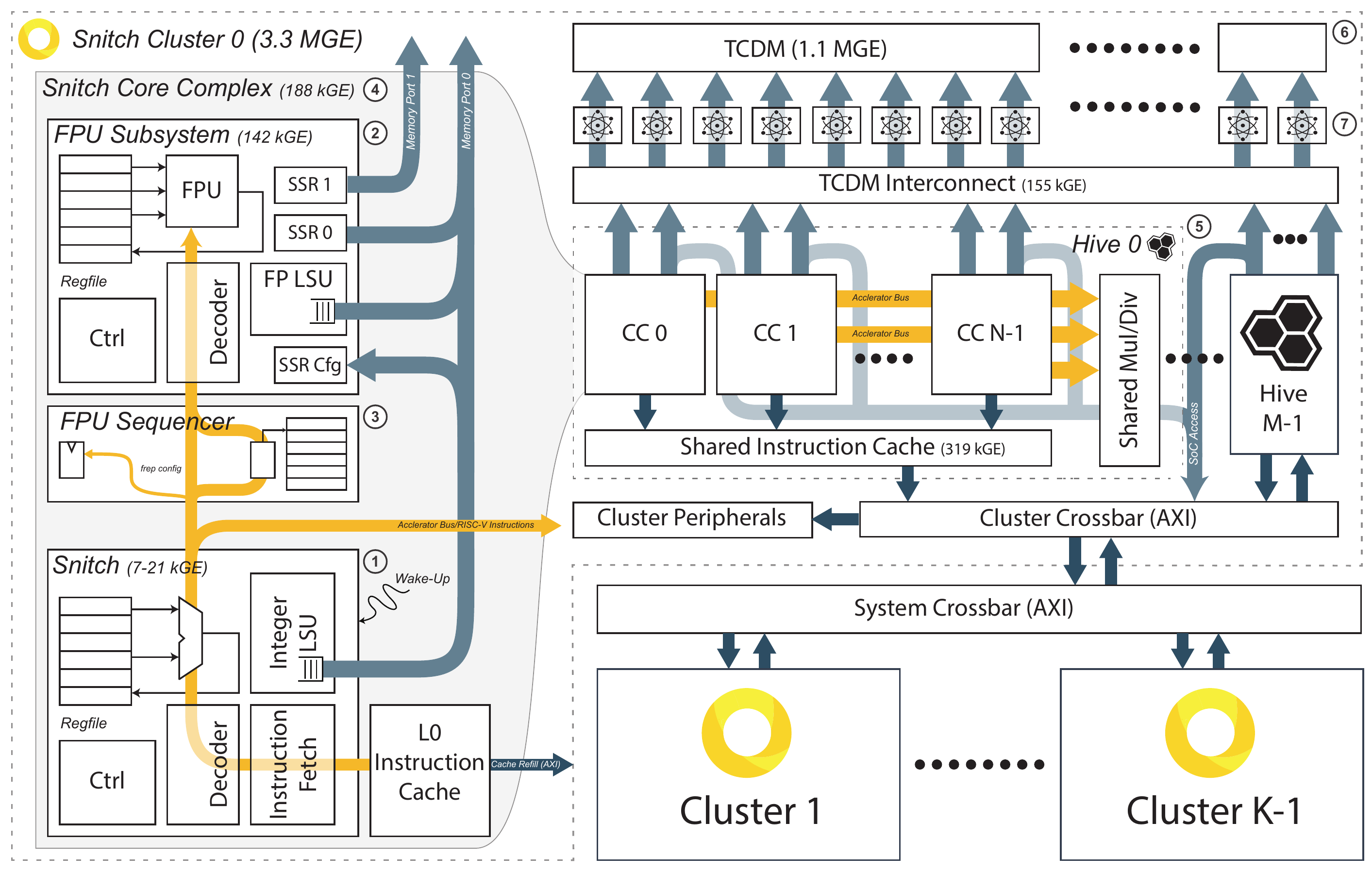}
    \caption{(4) Overview of an entire Snitch system. The smallest unit of repetition is a Snitch \gls{cc}. (1) It contains the integer core and the (2) FPU subsystem. (3) The \gls{fpu} sequencer, which is situated between the core and the \gls{fpss}, can be micro-coded to issue \fp instructions to the \gls{fpu} automatically. (5) The core is repeated $N$ times to form a Snitch Hive. Cores of a Hive share an integer multiply/divide unit and an L1 instruction cache. (6) $M$ Hives make up a Snitch Cluster that shares a \gls{tcdm}, a software-managed L1 cache. $K$ clusters are sharing last level memory via a crossbar. (7) Each \gls{tcdm} bank has a dedicated atomic unit that performs read-modify-write operations on its bank.}
    \label{fig:snitch_cluster_architecture}
\end{figure*}

\figref{fig:snitch_cluster_architecture} depicts the microarchitecture of the proposed system. The smallest unit of repetition is a Snitch \glsfirst{cc}. It contains the integer core and the \gls{fpu} subsystem. The core is repeated $N$ times to form a Snitch Hive. Cores of a Hive share an integer multiply/divide unit and an L1 instruction cache. $M$ Hives make up a Snitch Cluster that shares a \gls{tcdm} acting as a software-managed L1 cache. $K$ clusters share last level memory via a crossbar. All the parameters can be freely adjusted. For example, a Hive can just contain one core, therefore effectively making it a private multiplier and instruction cache. Similarly, a cluster can just contain one Hive with one core, making the \gls{tcdm} a private scratchpad memory.

\subsection{Snitch Core Complex}

The smallest unit of repetition is a Snitch \gls{cc}, see \figref{fig:snitch_cluster_architecture} (4). It contains an RV32IMAFD (RV32G) \riscv core and can be configured with or without support for the proposed \gls{isa} extensions. Depending on the technology and desired speed targets of the design, the offloading request, response, and the load/store interface to the \gls{tcdm} can be fully decoupled, increasing the design's clock frequency at the expense of increased latency of one cycle.

\subsubsection{Integer Core}
\label{subsec:integer_core}

The foundation of the system is an ultra-small (\SIrange{9}{20}{\kGE}), and energy-efficient 32\,bit integer \riscv compute unit, which we call \emph{Snitch} (\figref{fig:snitch_cluster_architecture} (1)). Snitch implements the entire (mandatory) integer base (RV32I). As its \glsfirst{rf} dominates the design of the \gls{cpu} implementation, we alternatively also support the embedded profile (E) as the other implementation choice. The embedded profile only provides 15 integer registers instead of 31. In addition, the \gls{rf} can either be implemented based on D-latches or D-flipflops. Each Snitch has a dedicated instruction fetch port, a data port with an independent \vtr~\cite{taylor2018basejump} decoupled request and response path, and a generic accelerator offloading interface. The accelerator interface has full support for offloading an entire 32\,bit \riscv instruction, and we re-use the same \riscv instruction encoding. This saves energy in the core's decoding logic as only a few bits need to be decoded to decide whether to offload an instruction or not. The interface has two independent decoupled channels. One for offloading an operation, up to three operands, and a back-channel for writing-back the result of the offloaded operation. In the presented design, we use the accelerator port to offload integer multiply/divide and floating-point instructions.

As our system's workload focuses on \fp computation Snitch was implemented with a minimal area footprint. The core is a single-stage, single-issue, in-order design. Integer instructions with all of their operands available (no data dependencies present) can be fetched, decoded, executed, and written back in the same cycle. We chose this design point to maximize energy efficiency and keep the design area at a minimum. The core keeps track of all 31 registers (the zero register is not writable, hence it does not need dedicated tracking) using a single bit in a scoreboard.  There are three classes of instructions that need special handling:

\paragraph{Integer instructions}

Most of the instructions contained in the \riscv I subset, such as integer arithmetic instructions, manipulation of \glspl{csr}, and control flow changes, can be executed in a single-cycle as soon as all operands are available. Integer multiply/divide instructions are part of the M subset and are offloaded to the (possibly) shared multiply/divide unit. There is no source of stalling as the \gls{alu} is fully combinational and executes its instruction in a single cycle. To foster the re-use of the \gls{alu}, it also performs comparison for branches, calculates \gls{csr} masks, and performs address calculations for load/store instructions.

\paragraph{Load/Store instructions}

Load/store instructions execute as soon as all operands are available, and the memory subsystem can process a new request. The data port of the core can exert back-pressure onto the load/store subsystem. Furthermore, the \gls{lsu} needs to keep track of issued load instructions and perform re-alignment and possible sign-extension. The core can have a configurable number of outstanding load instructions to the non-blocking memory hierarchy. Store instructions are considered fire-and-forget from a core perspective. The memory subsystem needs to maintain issue order as the core expects the arrival of load values in-order.

In addition to regular load and stores, the \gls{lsu} can also issue atomic memory operations and \gls{lrsc} as defined by the \riscv atomic memory operation specification. From a core perspective, the only difference is that the core also sends an atomic operation to the memory subsystem alongside the address and data. We provide additional signaling to accomplish that.

\paragraph{Accelerator/special function unit instructions}

Off-loaded instruction can execute as soon as all operands are available, and the accelerator interface can accept a new offloading request. We distinguish three types of instructions:
\begin{enumerate}
    \item Both destination and source operands are in the integer \gls{rf}, such as integer multiplication and division. Snitch's scoreboard keeps track of the destination operand.
    \item Source operands are in the integer \gls{rf}, and the receiving unit maintains the destination register. Such an example would be a move from integer to \fp \gls{rf}.
    \item Both operands are outside of the integer \gls{rf}, such as any \fp compute instruction (e.g., \gls{fma}).
\end{enumerate}

We offload floating-point instructions to the core-private \glsfirst{fpss} (\secref{subsec:fpu_subsystem}). As most of the \fp instructions operate on a separate \fp \gls{rf} we can easily decouple the \fp logic from the integer logic. The \riscv \gls{isa} specifies explicit move instructions from and to the \fp \gls{rf}, which makes this \gls{isa} particularly amenable for such an implementation. Decoupling the \gls{fpss} from the integer core makes it possible to alter and sequence \fp instructions into the \gls{fpss}. This is discussed in detail in~\secref{sec:fpu_sequencer}.

The second compelling use-case of the accelerator interface is to share expensive but, in our case, uncommonly used resources~\cite{gonzalez2010processor}. We provide a hardware implementation of the multiplication and division instructions for \riscv (M). This includes a fully pipelined 32\,bit multiplier, and a 32\,bit bit-serial integer divider with preliminary operand shifting for an early-out division — all cores of a Hive share such a hardware multiply/divide unit. Integer multiplications are two-cycle instructions while divisions are bit-serial and take up to 32 cycles in the worst case. By controlling the number of cores per Hive, the designer can adjust the sharing ratio. Sharing is independent of the functionality, and possibly many other resources can be shared, for example, a bit-manipulation \gls{alu}.

As the \gls{rf} only contains a single write-port, the three sources mentioned above contend over the single write port in a priority arbitrated fashion. Single-cycle instructions have priority over results from the \gls{lsu} over write-backs from the accelerator interface. That makes it possible to interleave results if an integer instruction does not need to write back, such as branch instructions, for example. Requests to the memory subsystem are only issued if there is space available to store the load result. Hence, cores cannot block each other with outstanding requests to the memory hierarchy. The integer core has priority on the register file to reduce the amount of logic necessary to retire a single-cycle instruction.

The Snitch integer core is formally verified against the \gls{isa} specification using the open-source \riscv formal framework~\cite{wolf2019formal}.

\subsubsection{FPU Subsystem}
\label{subsec:fpu_subsystem}

The \gls{fpss}, see \figref{fig:snitch_cluster_architecture} (2,3), bundles an \fpstd compliant \gls{fpu} with a 32$\times$64\,bit \gls{rf}. The \gls{fpss} has its own dedicated scoreboard where it keeps track of all registers in a similar fashion to the integer core. The \gls{fpu} is parameterizable in supported precision and operation latency \cite{mach2019fpu}. All \fp operations are fully pipelined (with possibly different pipeline depths). Operations without dependencies can be issued back to back. In addition to the \gls{fpu} it also contains a separate \gls{lsu} dedicated to loading and storing \fp data from/to the \fp \gls{rf}, the address calculation is performed in the integer core, which significantly reduces the area of the \gls{lsu}. Furthermore, the \gls{fpss} contains two \glspl{ssr} which map, upon activation through a \gls{csr} write, registers \verb+ft0+ and \verb+ft1+ to memory streams. The architecture of the streamers is depicted in~\figref{fig:streamer_architecture} and described in more detail in~\secref{sec:streamer}.

\subsection{Snitch Hive}

A Hive contains a configurable number of core complexes that share an instruction cache and a hardware multiply divide unit, see \figref{fig:snitch_cluster_architecture} (5).

Each core has a small, private, fully set-associative L0 instruction cache from which it can fetch instructions in a single cycle. A miss on the L0 cache generates a refill request upon the shared L1 instruction cache. If the cache-line is present, it is served from the data array of the L1 cache. If it also misses on the L1 cache, a refill request is generated and send to backing memory. Multiple requests to the same cache-line coalesce to a single refill request, which serves all pending requests. The L1 cache refills using an \gls{axi} burst-based protocol from the cluster crossbar.

The Snitch Hive serves another vital purpose: It provides a suitable boundary for separating physical design concerns. All signals crossing the design boundary are fully decoupled, and pipeline registers can be inserted to ease timing concerns on the boundaries of the design. The possibility to make a Hive the unit of repetition (a macro that is synthesized and placed and routed separately) allows for assembling larger clusters containing many more cores.

\subsection{Snitch Cluster}

One or more Hives make up a cluster, see \figref{fig:snitch_cluster_architecture} (6). Hives connect into the \gls{tcdm} crossbar that attaches to a banked shared memory, and the instruction refill port connects to the \gls{axi} cluster crossbar where it shares peripherals and communication to other clusters. The cluster crossbar provides both slave and master ports, which makes it possible to access the data of other clusters.

\subsubsection{Tightly Coupled Data Memory (TCDM)}

Core data requests are passed through an address decoder. Requests to a specific (configurable) memory range are routed towards the \gls{tcdm}, and all other requests are forwarded to the cluster crossbar. In its current implementation, the \gls{tcdm} crossbar is a fully connected, purely combinational interconnect. Other interconnect strategies can easily be implemented and will offer different scalability and conflict trade-offs. In order to reduce the effects of banking conflicts, we employ a banking factor of two, i.e., for each initiator port (two per core), we use two memory banks.

We resolve atomic memory operations and \gls{lrsc} issued by the core in a dedicated unit in front of each memory. The unit consists of a simple \gls{fsm} that performs the read-out of the operands from the underlying \gls{sram}. In the next cycle, it uses its local \gls{alu} to perform the required operations and finally saves the results in its memory. During the duration of an atomic operation, the unit blocks any access to the \gls{sram}.

\subsubsection{Cluster Peripherals}

The cluster peripherals are used by software to get information about the underlying hardware. Read-only registers provide information on \gls{tcdm} start and end address, number of cores per cluster, and \glspl{pmc} such as effective \gls{fpu} utilization, cycle count, \gls{tcdm} bank conflicts. Writable registers are a couple of scratch registers and a wake-up register, which triggers an \gls{ipi}.

\subsection{Stream Semantic Register (SSR)}
\label{sec:streamer}

\begin{figure}
    \centering
    \includegraphics[width=\linewidth]{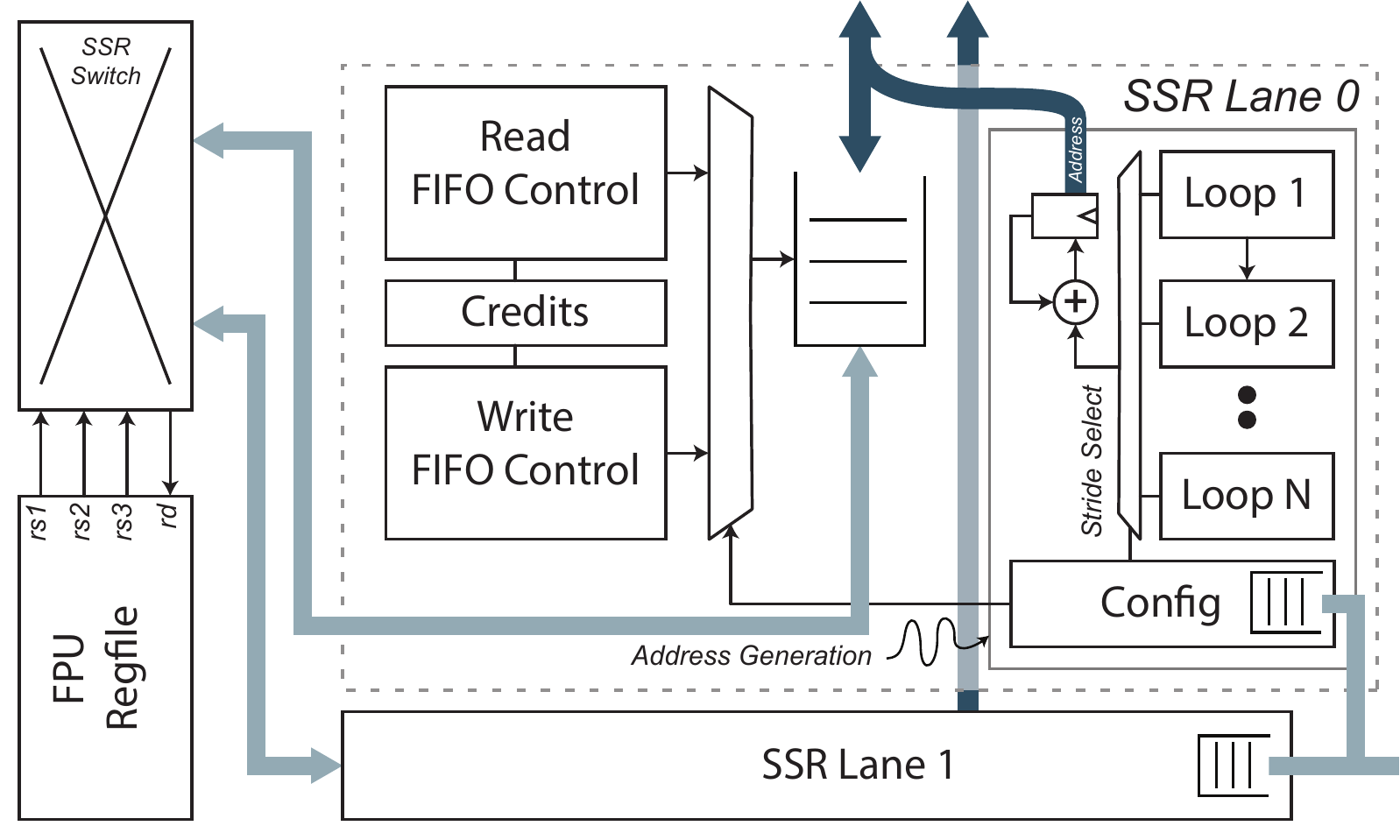}
    \caption{The \gls{ssr} hardware wraps around the \fp \gls{rf}. All three input and one output operands are mapped to two \gls{ssr} lanes. Each lane can be either configured as read or write and affine address calculation can be done with up to $N$ loop counters ($N$ is an implementation defined parameter). Requests are sent towards the memory hierarchy as soon as a valid configuration is in place. A credit-based queue hides the memory latency.}
    \label{fig:streamer_architecture}
\end{figure}

The \gls{ssr} extension was first proposed by Schuiki \emph{et~al.}~\cite{schuiki2018scalable, schuiki2019stream}. This hardware extension allows the programmer to configure up to two memory streams with an affine address pattern of dimension $N$. The dimension $N$ depends on the number of available loops (see~\figref{fig:streamer_architecture}) and can be parameterized. Streamers are configurable using memory-mapped \gls{io}. Each streamer is only configurable by the integer core controlling the \gls{fpss}. No other core can write the core-private configuration registers.

The \gls{ssr} module wraps logically around the \fp \gls{rf}. When activated by using a write to a \gls{csr}, operations on the \gls{rf} are intercepted iff the operands correspond to either \texttt{ft0} or \texttt{ft1} (which map to \gls{ssr} lane 0 or lane 1 respectively). The reads or writes are redirected towards an internal queue. The core communicates with the \gls{ssr} lane via a two-phase handshake. The core signals a valid request by pulling its read or write \emph{valid} signal high. In case data in the internal queue is available the respective \gls{ssr} lane signals \emph{readiness}. Finally, if the core decides to consume its register element it pulls its \emph{done} signal high.

For this work, we have extended the \gls{ssr}'s configuration scheme~\cite{schuiki2019stream} by adding shadow registers in which the core can already push the configuration of the next memory stream while the streaming is still in progress. This allows for overlapping loop-bound calculation with actual computation when using the \frep extension.

\subsection{FPU Sequence Buffer}
\label{sec:fpu_sequencer}

\begin{figure}
    \centering
    \includegraphics[width=\linewidth]{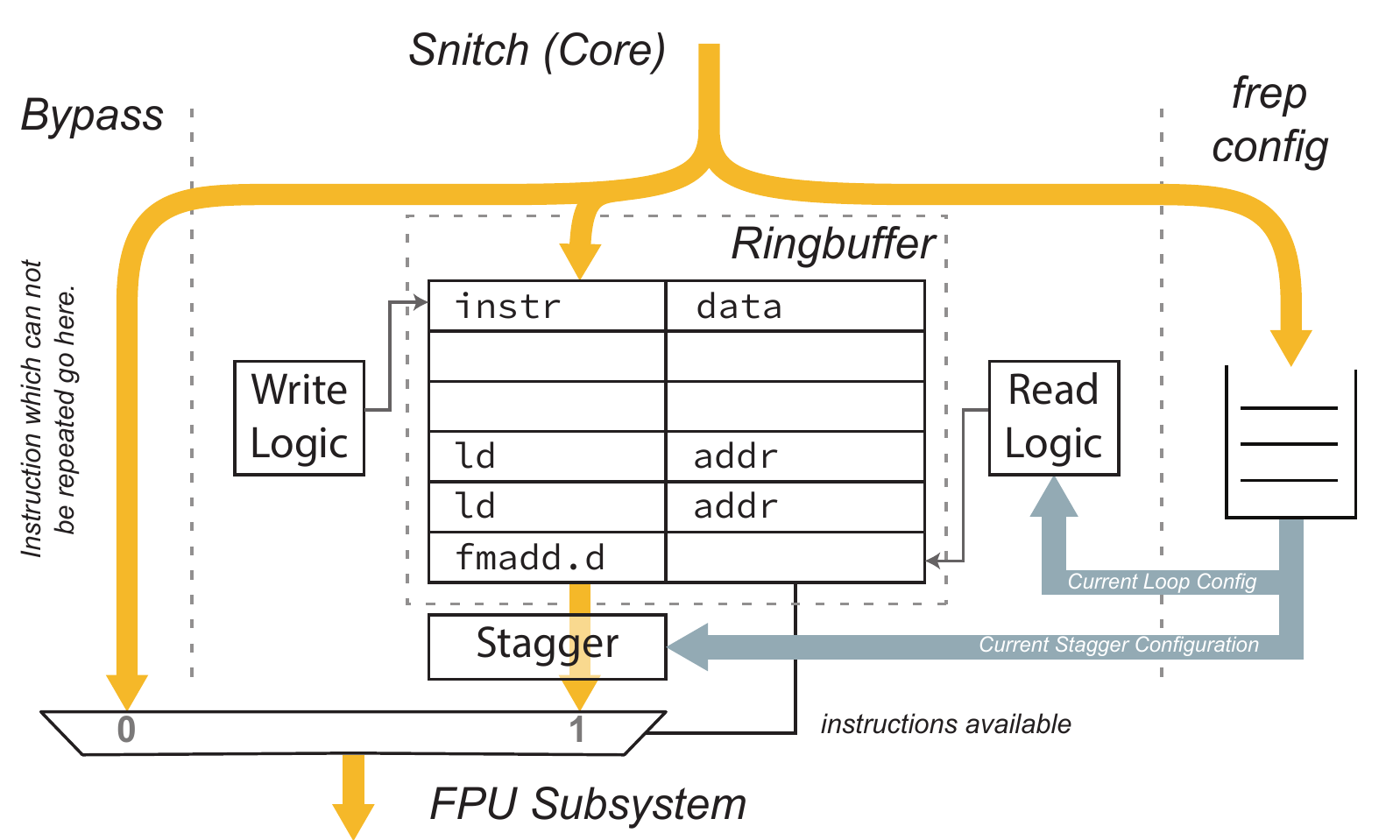}
    \caption{Microarchitecture of the \frep configurable \gls{fpu} sequence buffer. The core off-loads \fp instructions (top) to the \gls{fpss} (bottom). Depending on the instruction type (whether it is sequence-able), the instruction can use the bypass lane, be sequenced from the \gls{fpu} sequence buffer, or when an \frep instruction indicates another loop configuration request, it is saved into a configuration queue. The optional stagger stage can shift register operand names to avoid false dependency stalls and effectively provide a software-defined operand re-naming.}
    \label{fig:frep_architecture}
\end{figure}

\begin{figure}
    \centering
    \includegraphics[width=\linewidth]{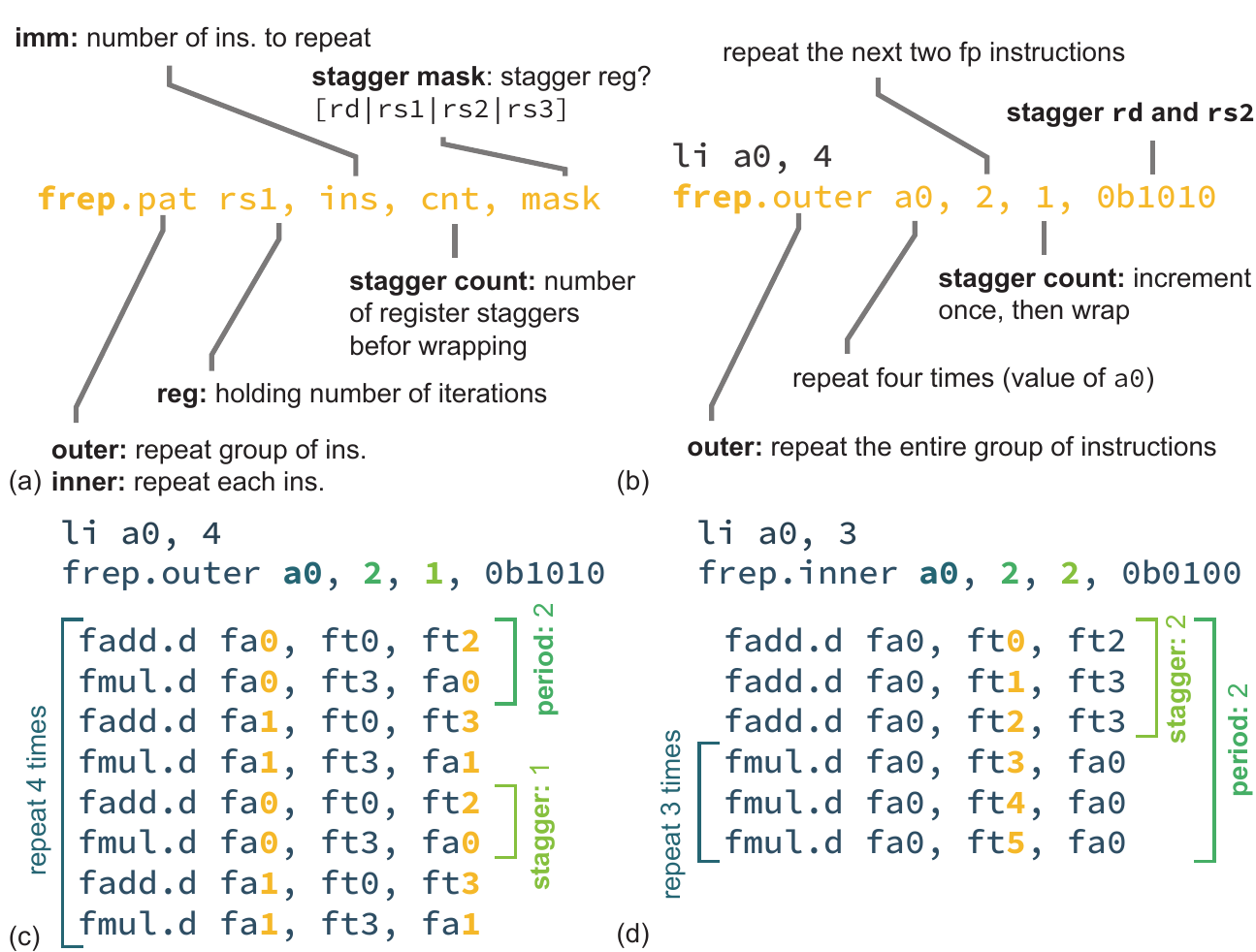}
    \caption{(a) Anatomy of the proposed \gls{frep} instruction. (b) An example usage of \gls{frep} sequencing the next two instructions a total of four times in an outer-loop configuration. (c) The corresponding instruction stream as sequenced to the \gls{fpss} including staggered registers (yellow bold face) and (d) another example sequencing two instructions for a total of three times in an inner-loop fashion and the resulting instruction stream with staggering highlighted.}
    \label{fig:frep_anatomy}
\end{figure}

The \gls{fpu} sequencer, depicted in~\figref{fig:frep_architecture}, is located at the off-loading interface between integer core and \gls{fpss}. It can be configured using the \frep instruction that provides the following information:
\begin{itemize}
    \item \verb+is_outer+: 1 bit indicating whether to repeat the whole kernel (consisting of \verb+max_inst+) or each instruction.
    \item \verb+max_inst+: 4-bit immediate (up to 16 values), indicates that the next \verb+max_inst+ should be sequenced.
    \item \verb+max_rep+: register identifier that holds the number of iterations (up to $2^{32}$ iterations)
    \item \verb+stagger_mask+: 4 bits for each operand (\verb+rs1 rs2 rs3 rd+). If the bit is set, the corresponding operand is staggered.
    \item \verb+stagger_count+: 3 bits, indicating for how many iterations the stagger should increment before it wraps again (up to $2^3=8$).
\end{itemize}
The \frep instruction marks the beginning of a \fp kernel which should be repeated, see \figref{fig:frep_anatomy} (a). It indicates how many subsequent instructions are stored in the sequence buffer, how often and how (operand staggering, repetition mode) each instruction is going to be repeated. To illustrate this we have given two examples in \figref{fig:frep_anatomy} (b, c, d). The first example sequences a block of two instructions a total of four times. The second example sequences two instructions three times. For this example, the sequencing mode is inner, meaning that each instruction is sequenced three times before the sequencer steps to the next instruction in the block.

A particular difficulty arises from the fact that, due to speed requirements, the \gls{fpu} is (heavily) pipelined, and \fp instructions take multiple cycles until their results become available for subsequent instructions. If the sequencer is going to sequence a short loop with data-dependencies amongst its operands, then the \gls{fpss} is going to stall because of data dependencies and therefore deteriorating performance, effective \gls{fpu} utilization, and energy efficiency. To mitigate the effects of stalling, the sequencer can change the register operands, indicated by a stagger mask, by adding a staggering count. \figref{fig:frep_anatomy} (c, d) demonstrates the sequencer's staggering capabilities. The first example (c) staggers the destination register, and the second source register a total of two times. The second example only staggers the first source register a total of 3 times.

\section{Programming}
\label{sec:programming_model}

\begin{figure}
    \centering
    \includegraphics[width=\linewidth]{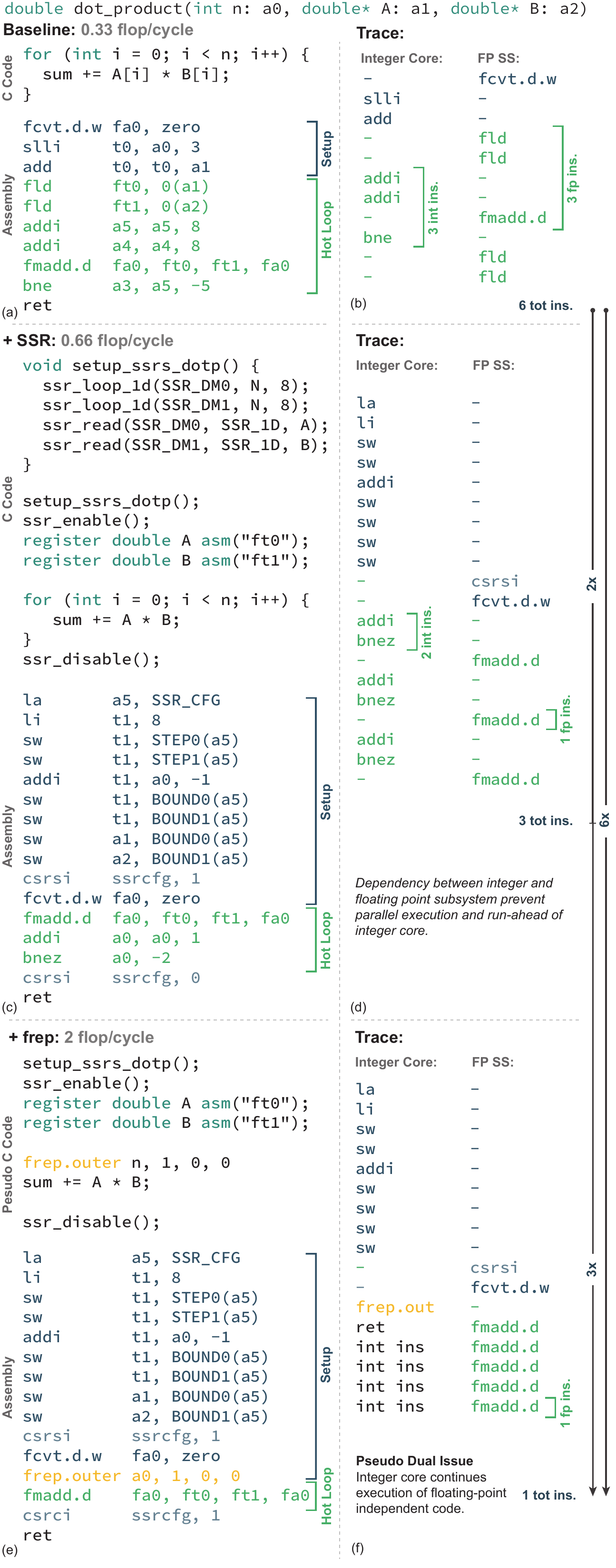}
    \caption{A dot product kernel in C and the corresponding \riscv assembly for all three extensions (a), (c), (e). Traces of each kernel are shown in (b), (d) and (f). Speed-ups of 2x and 6x for the proposed extensions. (f) also depicts the pseudo dual issue behavior.}
    \label{fig:dotp_assembly}
\end{figure}

\begin{figure}
    \centering
    \includegraphics[width=\linewidth]{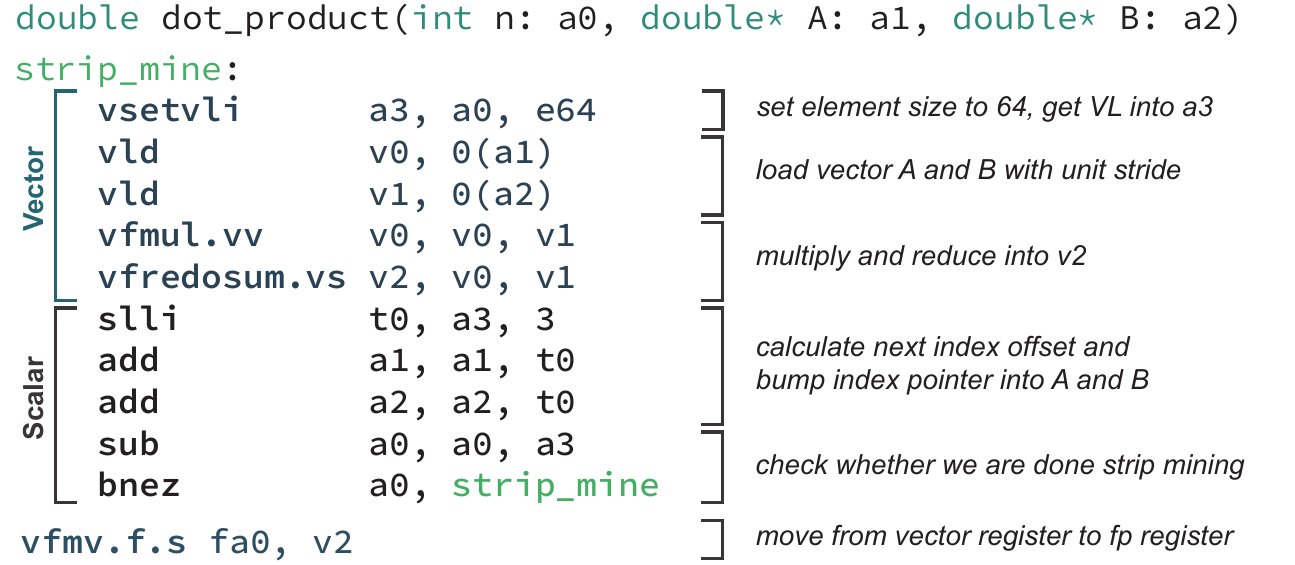}
    \caption{The same dot product kernel as in \figref{fig:dotp_assembly} in \riscv vector assembly~\cite{vector2020riscv}. The vector code is written independently of the \glsfirst{vl}, software needs to break the input problem size $n$ down to \gls{vl} in a strip mine loop. Of the ten instructions in the strip mine loop, five instructions are executed on the integer core while the other half is executed on the vector unit.}
    \label{fig:vector_assembly}
\end{figure}

Changing environments require a programmable system. To avoid overspecialization, we propose a system composed of many programmable and highly energy-efficient processing elements by leveraging widely applicable \gls{isa} extensions. At the foundation, the proposed system is a general-purpose \riscv-based multi-core system. The system has no private data caches but offers a fast, energy-efficient, and high-throughput software managed \gls{tcdm} as an alternative. It can be efficiently programmed using a \riscv toolchain, see \figref{fig:dotp_assembly}(a). The hardware provides atomic memory operations as defined by \riscv for efficient multi-core programs. The program operates on physical addresses with a minimal runtime.

The \gls{ssr} and \gls{frep} extension can be used with the provided header-only C library using an intrinsic-like style, similar to the \riscv vector intrinsics currently under development~\cite{bsc2020vector}. A set of, hand-tuned library routines can be used to exploit the proposed \gls{ssr} and \gls{frep} hardware extensions for optimal benefit of the proposed \gls{isa} extensions, similar to the cuBLAS or cuDNN libraries provided for Nvidia's \glspl{gpu}.
Furthermore, a first \gls{llvm} prototype shows that automatic code generation for \gls{ssr} setup is feasible~\cite{schuiki2019stream}.
\subsection{Stream Semantic Registers}

We provide a small, header-only, software library to program the \gls{ssr} efficiently. In particular, the programmer can decide the dimension of the stream and select the appropriate library function. For each dimension, the programmer needs to provide a stride, a bound, and a base address to configure the streamer. Finally a write to the \gls{ssr} \gls{csr} activates the stream semantic on register \verb+ft0+ and \verb+ft1+. After the streaming operation finishes, the same \gls{csr} is cleared to deactivate the extension. The whole programming sequence for an example kernel is depicted in \figref{fig:dotp_assembly}(c). On the example of the dot product kernel, we can see the speed-up of using the \gls{ssr} extension over the baseline implementation. The vanilla \riscv implementation executes a total of six instructions in its innermost loop, of which three are integer, and three are \fp instructions, see~\figref{fig:dotp_assembly}(b). The \gls{ssr}-enhanced version, on the other hand, elides all loads and only needs to track one loop counter to determine the loop termination condition. This saves three instructions and provides a 2x speed-up. The loop setup overhead is slightly higher, and a detailed analysis can be found in the original \gls{ssr} paper~\cite{schuiki2019stream}. For this system, we have enhanced the \gls{ssr} system to provide the programmer with shadow registers for the loop configuration. Therefore, the integer core can already set up the next loop iteration and store the configuration in the shadow registers while the current iteration is still in progress. When the current iteration finishes, the \gls{ssr} configuration logic automatically starts the iteration for the new configuration.

\subsection{\Gls{fpu} Sequencer}
\label{subsec:fp_sequencer}

The \verb+frep+ instruction configures the \gls{fpu} sequencer to automatically repeat and autonomously issue the next $n$ \fp instructions to the \gls{fpu}. This completely elides all loop instructions in the innermost loop iteration as the branch decision and loop counting is pushed to the sequencer hardware. For the dot product example, this only leaves one instruction in the innermost loop and provides a speed-up of $6\times$ compared to the baseline, and a $3\times$ improvement over the plain \gls{ssr} version of the kernel see \figref{fig:dotp_assembly}(f). As the \gls{fpu} sequencer frees the integer core of issuing instructions to the \gls{fpss}, it can continue executing integer instructions. This makes the core \emph{pseudo dual-issue}, see \figref{fig:dotp_assembly}(f). The pseudo-dual issue is a property of the decoupled design of \gls{fpss} and integer core: Both subsystems will execute as many instructions in parallel until they detect a blocking event such as a data movement from or to the \gls{fpss} and a dependent instruction.

For the same dot product kernel, we have also listed the corresponding \riscv vector assembly as a comparison point, see \figref{fig:vector_assembly}. Depending on the hardware's maximum \gls{vl} and the problem size, software needs to perform a strip mine loop over the input data. For each iteration, the \verb+setvl+ instruction saves the number of elements of subsequent vector instructions into its destination register. The integer core performs bookkeeping and pointer arithmetic for each iteration. Of the ten instructions of the strip mine loop, only five execute on the vector unit, of which only two perform arithmetic operations.

\subsubsection{Operand Staggering}

The complex \fp operations per\-for\-med by the \gls{fpu} require pipelining to achieve reasonable clock frequencies. Pipelining, on the other hand, increases the latency of \fp instructions, which makes it impossible for one \fp instruction to directly re-use the result of the previous instruction without stalling the pipeline. Depending on the speed target, we expect between two and six pipeline stages for \fp multiply-add. Therefore the next operation would need to wait for the same number of cycles until the operand becomes available. Some of these stalls can be hidden by executing independent \fp operations in the meantime. This technique requires partial unrolling of the kernel. To combine this efficiently with the \gls{frep} extension, we provide an option for the sequencer to stagger its operands. The staggering logic automatically increases the operand names of the issued instruction by one. The \verb+frep+ command takes an additional stagger mask and stagger count. The mask defines which register should be staggered. The mask contains one bit for all three source operands and the destination operand, four bits in total. If the corresponding bit is set, the \gls{fpu} sequencer increases the register name by one until the stagger count has been reached. Once the count is reached, the register name wraps again. The anatomy of the \verb+frep+ instruction including a sample trace with staggering enabled can be seen in \figref{fig:frep_anatomy} (a).

\section{Results}
\label{sec:results}
\begin{figure}
    \centering
    \includegraphics[width=\linewidth]{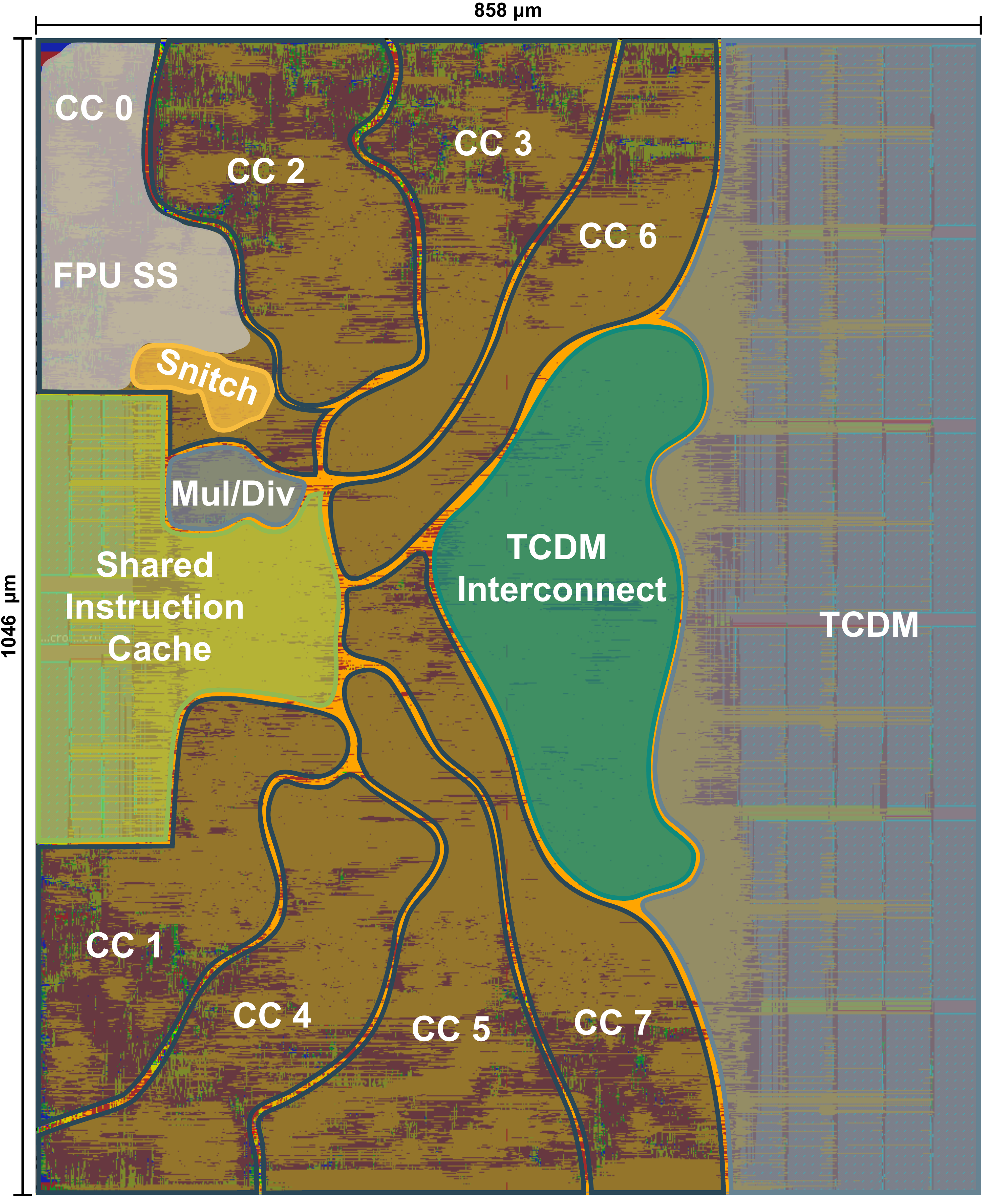}
    \caption{Placed and routed design of a Snitch Cluster. The cluster is configured to contain eight cores per Hive and one Hive per cluster. For \gls{cc} 0 we also highlighted the Snitch core and the \gls{fpss}. The configuration contains \num{32} banks of \gls{tcdm}, a total of \SI{128}{\kibi\byte} and \SI{8}{\kibi\byte} of instruction cache memory.}
    \label{fig:die_shot}
\end{figure}

We have synthesized, placed and routed an eight core configuration with two hives (each with four cores), \SI{128}{\kibi\byte} of \gls{tcdm}, and \SI{8}{\kibi\byte} of instruction cache using the \designcompiler and \innovus in a modern \gf technology. The floorplan of this cluster is depicted in \figref{fig:die_shot}. For the synthesis we have constrained the design to close timing at \SI{1}{\giga\hertz} in worst case conditions (SSG\footnote{p-channel \gls{mosfet} globally slow, n-channel \gls{mosfet} globally slow}, \SI{0.72}{\volt}, \SI{-40}{\degreeCelsius}). The subsequent place and route step was constrained to \SI{0.7}{\giga\hertz}. Sign-off \gls{sta} using \primetime showed that the design runs at \SI{755}{\mega\hertz} in worst case conditions and \SI{1.06}{\giga\hertz} in typical conditions (TT\footnote{p-channel \gls{mosfet} typical, n-channel \gls{mosfet} typical}, \SI{0.8}{\volt}, \SI{25}{\degreeCelsius}).

\subsection{Microkernels}
\label{subsec:microkernels}

To evaluate the performance, power, and energy-efficiency of the architecture, we have implemented a set of different data-oblivious parallel benchmarks, where the control flow only depends on a constant number of program parameters. We selected four complementary kernels:
\begin{itemize}
    \item Dot product: A simple dot product implementation that calculates the scalar product of two arrays of length $n$. Included because it is a fundamental vector-vector operation (\gls{blas} 2).
    \item ReLU: This kernel applies a \gls{relu} to the elements of an array of length $n$. The kernel is often used as an activation function for neural networks ($n$ \gls{blas} $1^n$).
    \item Matrix multiplication using the dot product method: A chunked implementation of matrix multiplication of size $n\times n$. A highly relevant kernel for the machine learning domain (\gls{blas} 3). The output matrix is chunked across the cores.
    \item \Gls{fft}: Implementation of a parallel \gls{fft} algorithm of size $n$. Included to show the versatility of the tightly coupled core and the proposed extensions. The \gls{fft} is based on Cooley–Tukey's algorithm.
    \item AXPY ($a\cdot\vec{x} + \vec{b}$) on vectors of length $n$: Included as a memory-bound kernel. As the benchmarked system only provides two \glspl{ssr}, the core needs to perform the store operation, which prevents any speed-ups from \gls{frep}. Furthermore, the kernel is memory-bound as it requires three memory accesses per two \fp operations but each core can only sustain two memory operations through its two ports in the \gls{tcdm} interconnect (\gls{blas} 1).
    \item kNN: This algorithm performs a point-wise Euclidean distance calculation between all points ($n$) in the system and a sample. In a second sorting step, the $k$ closest points are returned as classification results. The \gls{ssr}+\gls{frep} can significantly speed-up the Euclidean distance calculation. However, the dominant factor of the over-all runtime is the sorting step, which can not easily be accelerated using \gls{ssr} and \gls{frep}. To provide maximum insight into the achievable improvement, we focused our measurements on the distance calculation. Parallelization is achieved by distributing the sampling step amongst all cores.
    \item Monte Carlo method approximating $\pi$ in $n$ steps. The integer core generates random numbers while the floating-point subsystem evaluates the function to be integrated. \gls{ssr} and \gls{frep} make for an exciting application since the pseudo-dual issue allows the two tasks to entirely overlap and execute in parallel on the integer core and floating-point subsystem, respectively. Interestingly, we see a slight drop in speed-up in the pure \gls{ssr} case because the problem needs to be re-formulated for \gls{ssr} usage, which in turn exhibits an adversarial instruction pattern in the \gls{fpss} (many dependent \fp instructions).
    \item 2D Convolution on a $32 \times 32$ image with a $7 \times 7$ kernel (kernel size is from the first layer of Google LeNet, the input image size has been truncated to reduce the problem runtime): A highly relevant workload for the machine-learning and data-science domain. The high data-reuse and affine access pattern make it an ideal candidate for enhancement with \glspl{ssr} and \gls{frep}.
\end{itemize}
For each kernel we provide a baseline C implementation\footnote{riscv32-unknown-elf-gcc (GCC) 7.2.0 -03} (without auto-vectorization or special intrinsics), an implementation which makes use of \glspl{ssr} and one which combines \glspl{ssr} and \gls{frep}. We have made sure (partially by using inline-assembly) that the generated baseline code is optimal and executes well on the Snitch core. Speed-ups are measured in a cycle-accurate \gls{rtl} simulation, similarly power estimations are measured in a post-layout simulation. All the kernels input and output data set sizes are chosen so that they fit into the \gls{tcdm} to avoid measuring effects of the cluster-external memory hierarchy.

\subsection{Single-Core}
\subsubsection{Performance}

\setlength{\tabcolsep}{3pt}
\renewcommand{\arraystretch}{1.2}
\begin{table}
    \begin{threeparttable}
        \caption{Single and multi-core utilization of the \gls{fpu}, the \gls{fpss}, the integer core, and total IPC for all benchmarks. A high baseline \gls{ipc} ensures a fair comparison with the proposed \gls{isa} extensions.}
        \begin{tabularx}{\linewidth}{@{}Xrrrrcrrrr@{}}
            \toprule
                                   & \multicolumn{9}{c}{Utilization}                                                                                                                                               \\
            \cmidrule{2-10}
                                   & \multicolumn{4}{c}{Single-Core} &      & \multicolumn{4}{c}{Multi-Core (8 Cores)}                                                                                             \\
            \cmidrule{2-5} \cmidrule{7-10}
            Kernel                 & FPU                             & FPSS & Snitch                                   & IPC           &  & FPU                   & FPSS                  & Snitch & IPC           \\
            \midrule
            Dot Pr. $256$          & 0.17                            & 0.50 & 0.50                                     & 1.00          &  & 0.20                  & 0.58                  & 0.22   & 0.80          \\
            + SSR                  & 0.61                            & 0.63 & 0.35                                     & 0.98          &  & 0.35                  & 0.38                  & 0.32   & 0.69          \\
            + SSR + \gls{frep}     & 0.87                            & 0.89 & 0.06                                     & 0.96          &  & 0.35                  & 0.41                  & 0.18   & 0.59          \\ \midrule
            Dot Pr. $4096$         & 0.25                            & 0.75 & 0.25                                     & 1.00          &  & 0.24                  & 0.70                  & 0.24   & 0.94          \\
            + SSR                  & 0.66                            & 0.66 & 0.34                                     & 1.00          &  & 0.57                  & 0.58                  & 0.32   & 0.90          \\
            + SSR + \gls{frep}     & 0.98                            & 0.99 & 0.01                                     & 0.99          &  & 0.72                  & 0.74                  & 0.05   & 0.79          \\ \midrule
            ReLU                   & 0.14                            & 0.42 & 0.57                                     & 1.00          &  & 0.13                  & 0.37                  & 0.53   & 0.90          \\
            + SSR                  & 0.32                            & 0.32 & 0.67                                     & 0.99          &  & 0.23                  & 0.23                  & 0.56   & 0.79          \\
            + SSR + \gls{frep}     & 0.88                            & 0.89 & 0.07                                     & 0.96          &  & 0.36                  & 0.36                  & 0.23   & 0.62          \\ \midrule
            DGEMM $16^2$           & 0.19                            & 0.58 & 0.17                                     & 0.75          &  & 0.17                  & 0.51                  & 0.15   & 0.66          \\
            + SSR                  & 0.23                            & 0.26 & 0.53                                     & 0.80          &  & 0.20                  & 0.23                  & 0.49   & 0.72          \\
            + SSR + \gls{frep}     & 0.86                            & 0.97 & 0.07                                     & \tnote{*}1.04 &  & 0.63                  & 0.71                  & 0.13   & 0.84          \\ \midrule
            DGEMM $32^2$           & 0.24                            & 0.26 & 0.52                                     & 0.77          &  & 0.24                  & 0.26                  & 0.51   & 0.77          \\
            + SSR                  & 0.24                            & 0.26 & 0.52                                     & 0.77          &  & 0.24                  & 0.26                  & 0.51   & 0.77          \\
            + SSR + \gls{frep}     & 0.93                            & 0.99 & 0.03                                     & \tnote{*}1.02 &  & 0.85                  & 0.90                  & 0.04   & 0.94          \\ \midrule
            \gls{fft}              & 0.36                            & 0.49 & 0.23                                     & 0.72          &  & 0.26                  & 0.35                  & 0.23   & 0.58          \\
            + SSR                  & 0.54                            & 0.58 & 0.32                                     & 0.90          &  & \tnote{$\dagger$}0.21 & \tnote{$\dagger$}0.23 & 0.41   & 0.65          \\
            + SSR + \gls{frep}     & 0.57                            & 0.62 & 0.19                                     & 0.81          &  & \tnote{$\dagger$}0.24 & \tnote{$\dagger$}0.27 & 0.42   & 0.69          \\ \midrule
            AXPY\tnote{$\ddagger$} & 0.19                            & 0.77 & 0.20                                     & 0.97          &  & 0.14                  & 0.63                  & 0.19   & 0.82          \\
            + SSR                  & 0.34                            & 0.67 & 0.27                                     & 0.95          &  & 0.23                  & 0.47                  & 0.30   & 0.77          \\ \midrule
            2D Conv.               & 0.14                            & 0.43 & 0.57                                     & 1.00          &  & 0.14                  & 0.42                  & 0.58   & 1.00          \\
            + SSR                  & 0.60                            & 0.60 & 0.39                                     & 0.99          &  & 0.60                  & 0.61                  & 0.39   & 0.99          \\
            + SSR + \gls{frep}     & 0.97                            & 0.99 & 0.04                                     & \tnote{*}1.03 &  & 0.91                  & 0.93                  & 0.04   & 0.97          \\ \midrule
            kNN                    & 0.15                            & 0.31 & 0.40                                     & 0.70          &  & 0.14                  & 0.31                  & 0.40   & 0.70          \\
            + SSR                  & 0.30                            & 0.30 & 0.64                                     & 0.95          &  & 0.30                  & 0.31                  & 0.66   & 0.97          \\
            + SSR + \gls{frep}     & 0.35                            & 0.36 & 0.76                                     & \tnote{*}1.13 &  & 0.35                  & 0.37                  & 0.79   & \tnote{*}1.16 \\ \midrule
            Monte Carlo            & 0.14                            & 0.18 & 0.59                                     & 0.77          &  & 0.13                  & 0.16                  & 0.54   & 0.70          \\
            + SSR                  & 0.15                            & 0.21 & 0.61                                     & 0.82          &  & 0.14                  & 0.20                  & 0.57   & 0.77          \\
            + SSR + \gls{frep}     & 0.22                            & 0.22 & 0.90                                     & \tnote{*}1.12 &  & 0.20                  & 0.20                  & 0.82   & \tnote{*}1.02 \\
            \bottomrule
        \end{tabularx}
        \begin{tablenotes}
            \item [*] \emph{Pseudo-dual issue} behavior with a cumulative \gls{ipc} higher than one \\
            \item [$\dagger$] Reduction of \gls{fpu} utilization because of \gls{ssr} setup and frequent re-synchronization between \gls{fft} stages. We still show a speed-up of $2.8\times$ (see \figref{fig:multi_core_speed_up}) \\
            \item [$\ddagger$]AXPY can not be enhanced using \gls{frep} because the current architecture provides only two streamers. For the AXPY kernel three streamers would be needed.
        \end{tablenotes}
        \label{tab:utilization}
    \end{threeparttable}
\end{table}

\begin{figure}
    \centering
    \includegraphics[width=\linewidth]{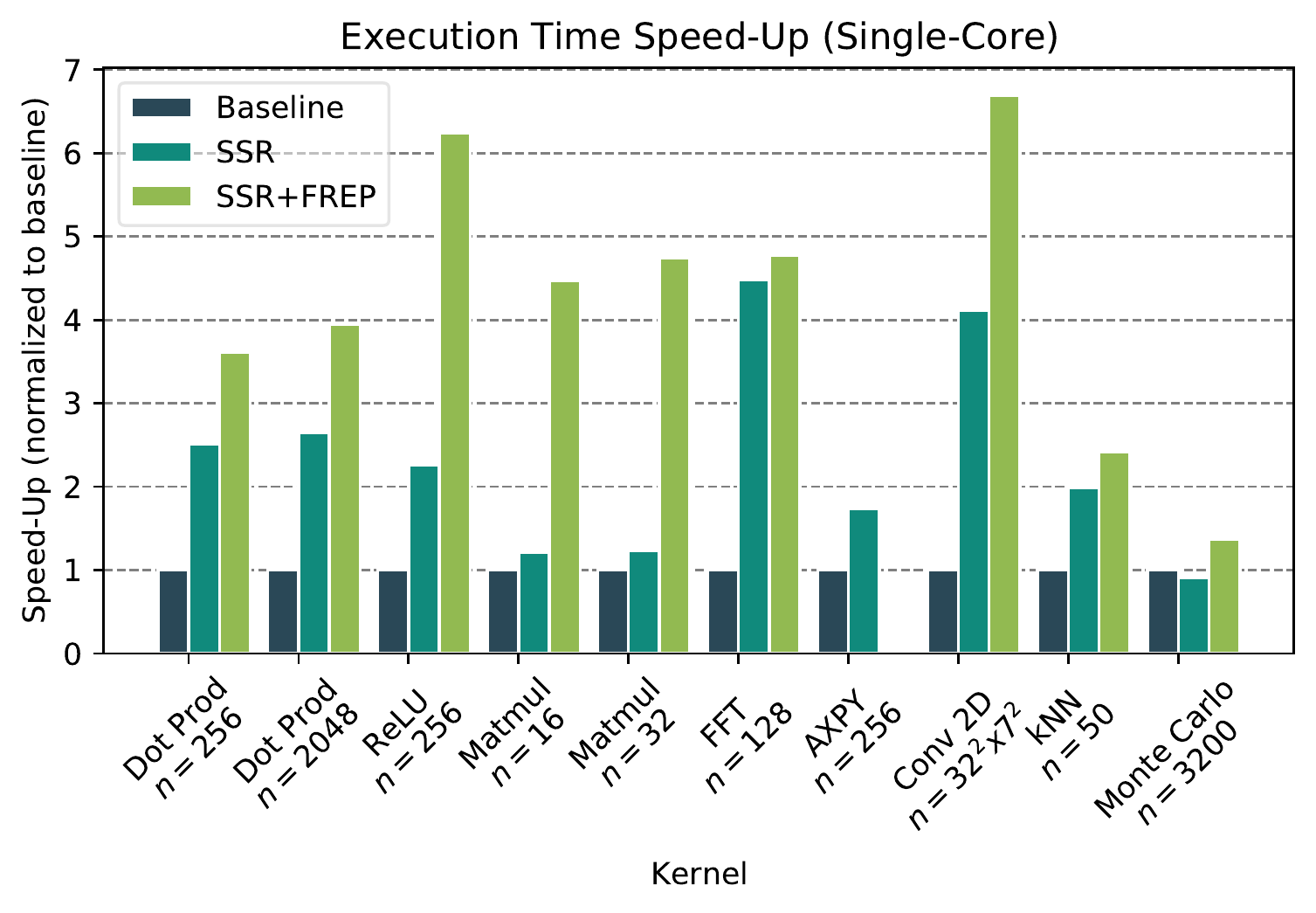}
    \caption{Single-core speed-up reported for each microkernel and enabled extension. By using our proposed \gls{ssr} and \gls{frep} extensions can achieve speed-ups from $1.7\times$ to over $6\times$ on selected benchmarks.}
    \label{fig:single_core_speed_up}
\end{figure}

The single-pipeline stage of the core lets it achieve a very high \gls{ipc} of close to one for most of the kernels. The only effective source of stalls comes from the memory interface if there is a load-use dependency present or when the load result contends for the single write port of the core's \gls{rf}. The proposed \gls{isa} extensions, \gls{ssr}, and \gls{frep} reduce the number of explicit load and store instructions as well as the branching overhead. For above-mentioned microkernels we can report single-core speed-ups of over 6x in \figref{fig:single_core_speed_up} on certain benchmarks. The single-core case presents an idealized execution environment as there is no contention on the shared \gls{tcdm}. We observe interesting effects: The matrix multiplication, 2D convolution,  kNN distance calculation, and the Monte Carlo benchmark achieve an \gls{ipc} of more than one by overlapping the computation of one block with the \gls{ssr} setup and integer instructions of the next block.

In \tabref{tab:utilization} we are tracking four metrics:
\begin{enumerate}
    \item \gls{fpu} utilization: The total number of arithmetic \fp instructions executed. We consider (fused) arithmetic operations, casts, and comparison instructions as \fp operations.
    \item \gls{fpss} utilization: Includes all instructions that are off-loaded to the \gls{fpss}. This counts all \fp instructions as well as \fp loads and stores.
    \item Snitch utilization: Contains all instructions that are not offloaded to the \gls{fpss}.
    \item Total \gls{ipc}: Snitch utilization and \gls{fpss} utilization result in the total \gls{ipc}. For the baseline case, this metric is interesting as due to the single pipeline stage and the tightly coupled memory subsystem we achieve an \gls{ipc} of one for almost every kernel in the single-core case. For the multi-core system, contentions on the memory interface slightly limit the attainable \gls{ipc}. This ensures a fair baseline for further evaluating our \gls{isa} extensions. The reported \gls{ipc} for the \gls{frep} enhanced kernels includes the \gls{frep} generated instructions.
\end{enumerate}

The single-issue nature of the baseline core limits the maximum achievable \gls{fpu} utilization as we need to explicitly move data from memory into the core's register file. This ranges from \num{0.14} to \num{0.36} depending on the benchmark. We can see a very high core utilization as the integer core is supplying the \gls{fpu} with instructions.

The introduction of \gls{ssr} relaxes these constraints as we are translating all loads and stores into implicitly encoded register reads. We can see a positive effect on execution time as we are not using an issue slot (cycle) of the integer core to issue load(s)/store(s). We can still see that the integer core is busy issuing arithmetic \fp instructions to the \gls{fpu} by observing a high Snitch utilization.

Finally, with the introduction of \gls{frep}, we significantly reduce the pressure on the integer core. The integer core only issues the \fp operations once into the \frep buffer from which it is being sequenced multiple times to the \gls{fpss}. We can observe a very low integer core utilization of somewhere between \numrange{0.03}{0.24}. As we free the integer core from issuing \fp instructions on every cycle, we can easily keep the \gls{fpu} busy. This results in a very high \gls{fpu} utilization of \numrange{0.57}{0.93}. A high \gls{fpu} utilization, in turn, means high energy efficiency. For the single-core case we can see an improvement in speed-up (see~\figref{fig:single_core_speed_up}) and \gls{fpu} utilization for all microkernels. The \gls{fft} benchmark shows a reduction in \gls{ipc} as more frequent \gls{ssr} set-up and load-use dependencies insert stall cycles which result in pipeline bubbles.

\subsubsection{Area}
\begin{figure}
    \centering
    \includegraphics[width=\linewidth]{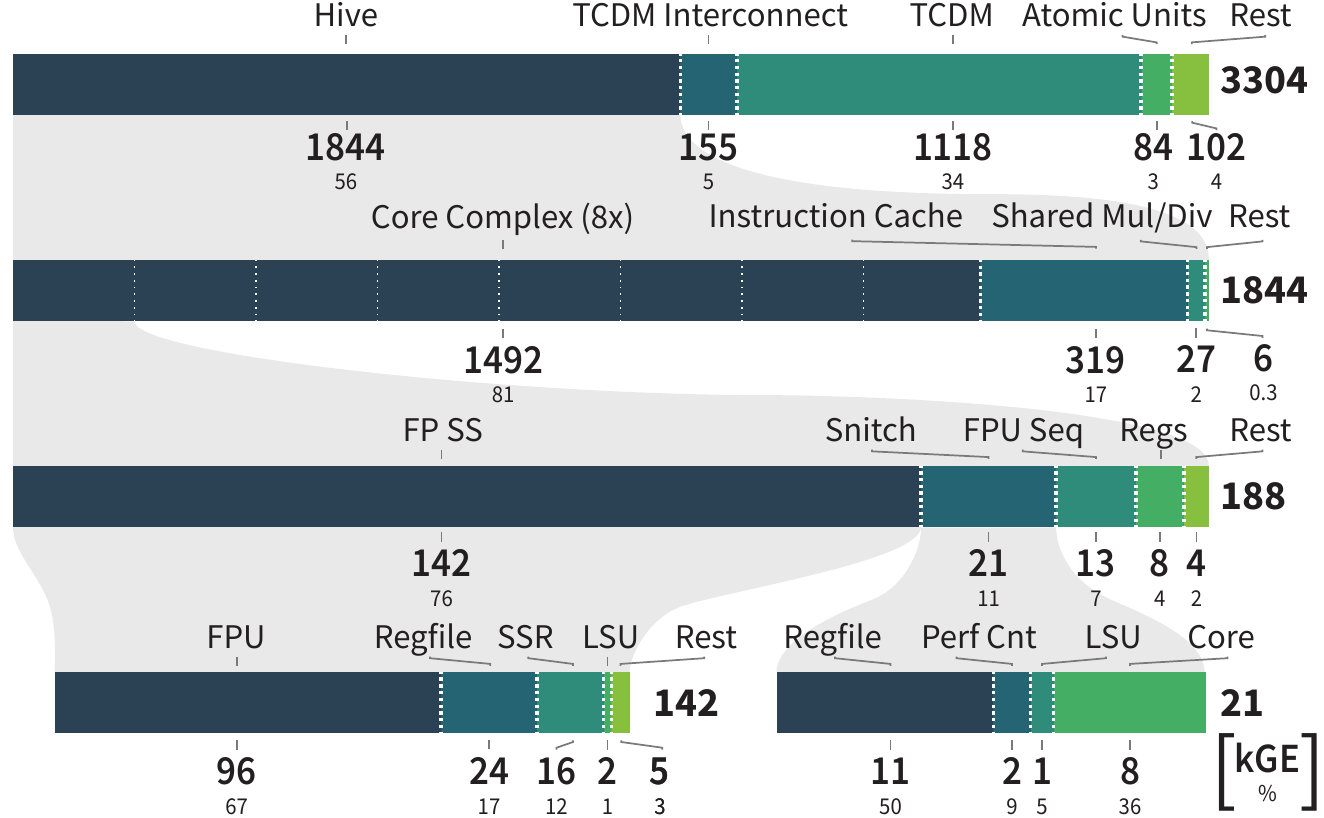}
    \caption{Hierarchical area distribution of the Snitch cluster. The entire cluster has a size of approximately \SI{3.3}{\mega\gateequivalent}. \SI{34}{\percent} of the area is occupied by the \gls{tcdm}. The instruction cache makes up for \SI{10}{\percent} of the cluster's area. Of each \gls{cc} the \gls{fpss} accounts for \SI{76}{\percent} while the integer core only accounts for \SI{11}{\percent} of the \gls{cc}'s area. In total all integer cores occupy only \SI{5}{\percent} of the cluster's total area while the \glspl{fpu} make up for over \SI{23}{\percent} of the total cluster area. The Snitch core has been configured with RV32I and a FF-based \gls{rf} and \glspl{pmc}. See \figref{fig:snitch_cluster_architecture} for an overview of the system's main components.}
    \label{fig:area_bowtruckle}
\end{figure}
\begin{figure}
    \centering
    \includegraphics[width=\linewidth]{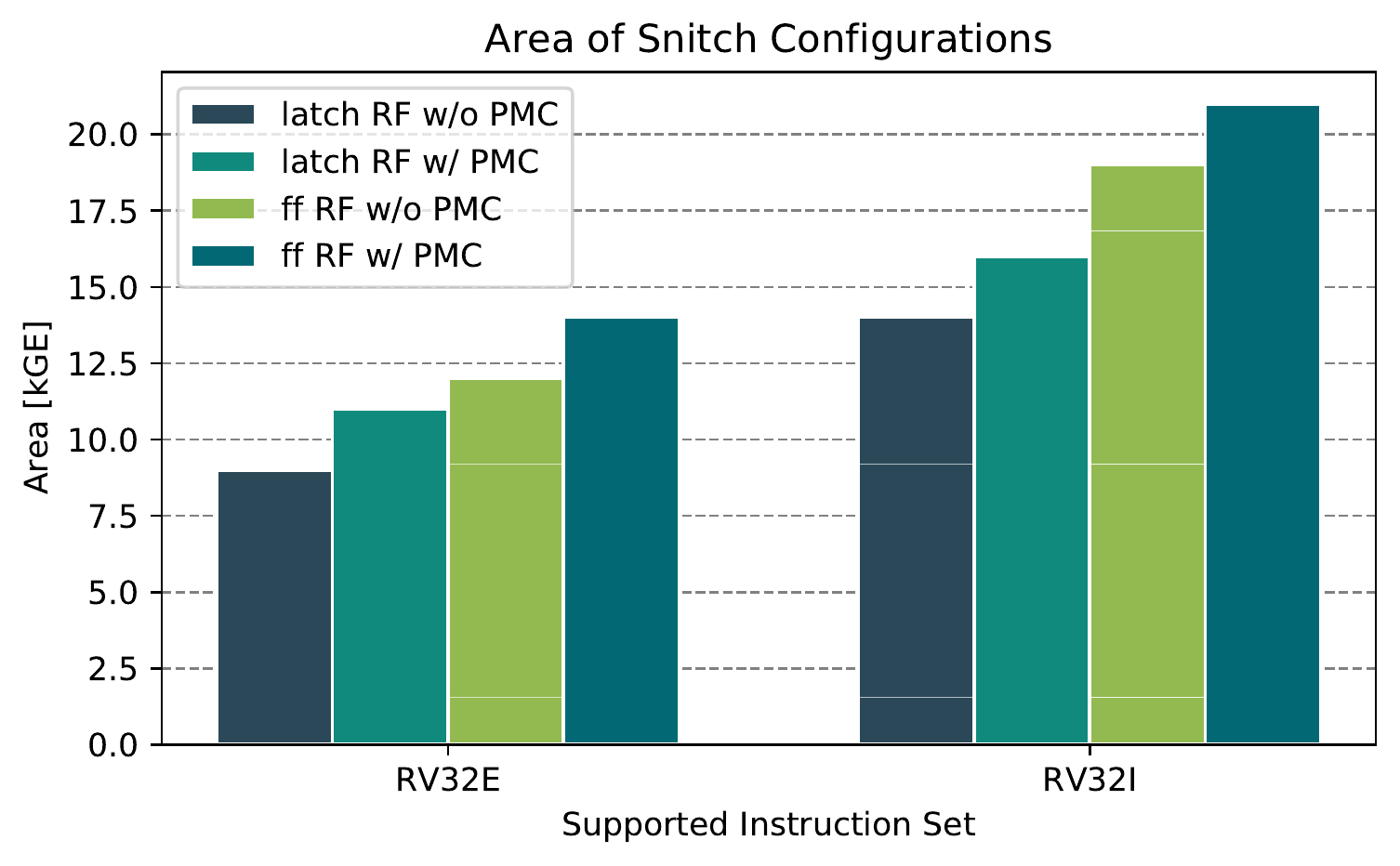}
    \caption{Area of different integer core configurations. We provide choice of the \gls{isa} variant, of the \gls{rf} and inclusion of \gls{pmc}.}
    \label{fig:area_snitch}
\end{figure}
The integer core \gls{isa} is configurable to either be RV32I or RV32E. Both support the same instructions but differ in the size of the \gls{rf}. While RV32I comes with 32 general purpose integer register, RV32E only provides 16. As the CPU design is heavily dominated by the \gls{rf} (see \figref{fig:area_bowtruckle}) this design choice has a significant influence on the core's area. Furthermore, as mentioned in \secref{subsec:integer_core} we provide a latch-based and a FF-based \gls{rf} implementation. The first being \SI{50}{\percent} smaller in area while the latter can be used if latches are not available in the standard-cell library. Moreover, \glspl{pmc} can be enabled separately which adds approximately \SI{2}{\kGE} in area. Altogether this makes the integer core configurable from \SI{9}{\kGE} (RV32E, latch-based \gls{rf} without \gls{pmc}) up to \SI{21}{\kGE} (RV32I, flip-flop-based \gls{rf} with \gls{pmc}), see \figref{fig:area_snitch}.
The \gls{ssr} hardware consumes \SI{16}{\kGE} to implement address generation and control logic as well as load data buffering. This puts it at \SI{12}{\percent} of the \gls{fpss} and \SI{8.5}{\percent} of the \gls{cc}. The \gls{frep} extension, configured with \num{16} entries, takes up \SI{13}{\kGE} which is \SI{7}{\percent} of the \gls{fpss}'s area and \SI{3.2}{\percent} of the overall \gls{soc} (a total of \SIrange{38}{50}{\kilo\gateequivalent} for the \gls{cc}).

\subsection{Multi-Core}
\label{subsec:multi_core}
\begin{figure}
    \centering
    \includegraphics[width=\linewidth]{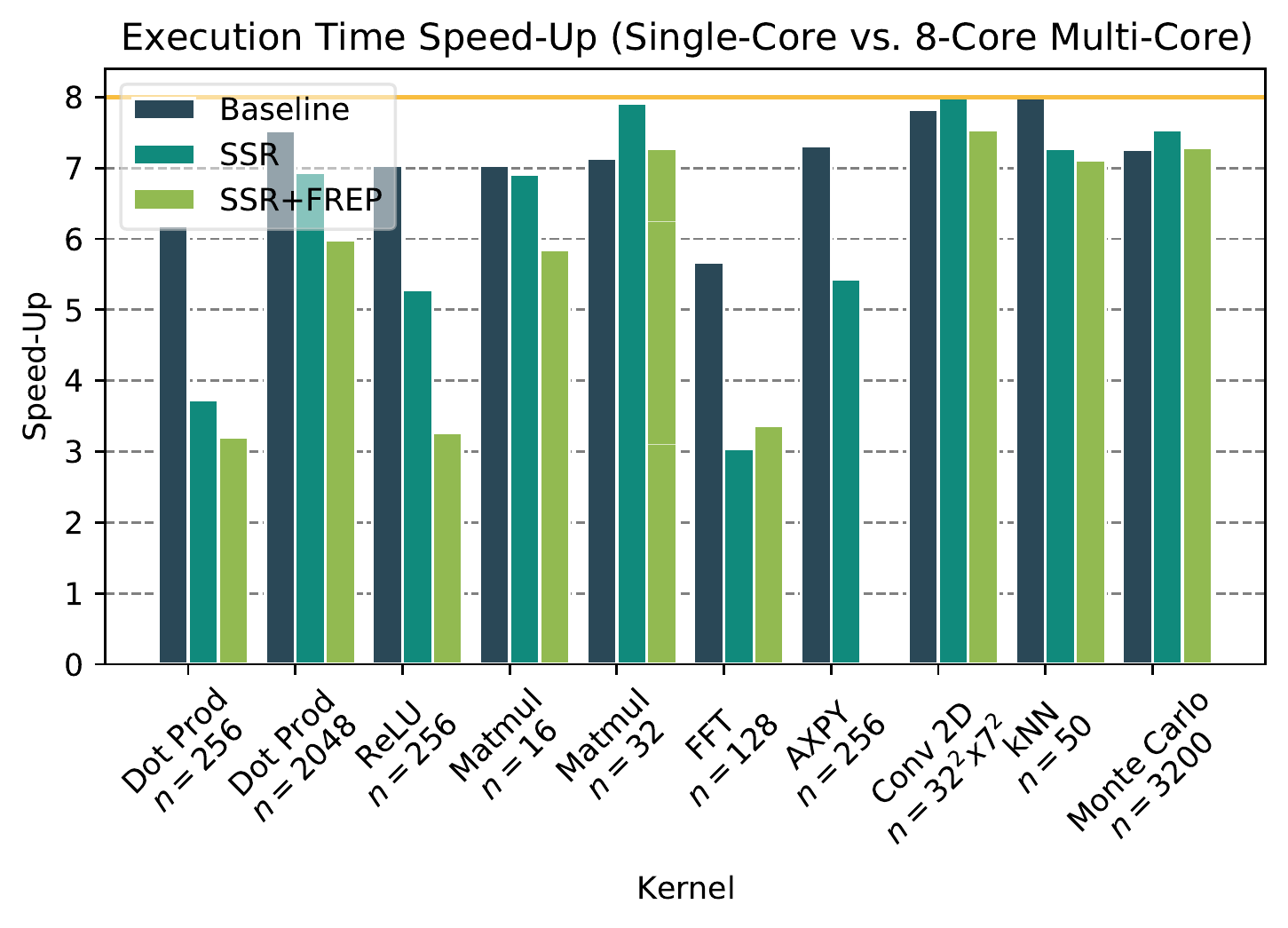}
    \caption{Single-core vs. an octa-core cluster speed-ups. Ideal speed-ups of eight are achieved for the pure \gls{ssr} 2D convolution and the kNN baseline. Very high multi-core speed-ups are measured for matrix multiplication, 2D convolution, kNN, and Monte Carlo methods. The \gls{fft}, dot product and AXPY show less speed-ups as (mostly due to the small problem size) the reduction and synchronization of all cores have a stronger impact on the runtime.}
    \label{fig:multi_core_vs_single_core_speed_up}
\end{figure}
\begin{figure}
    \centering
    \includegraphics[width=\linewidth]{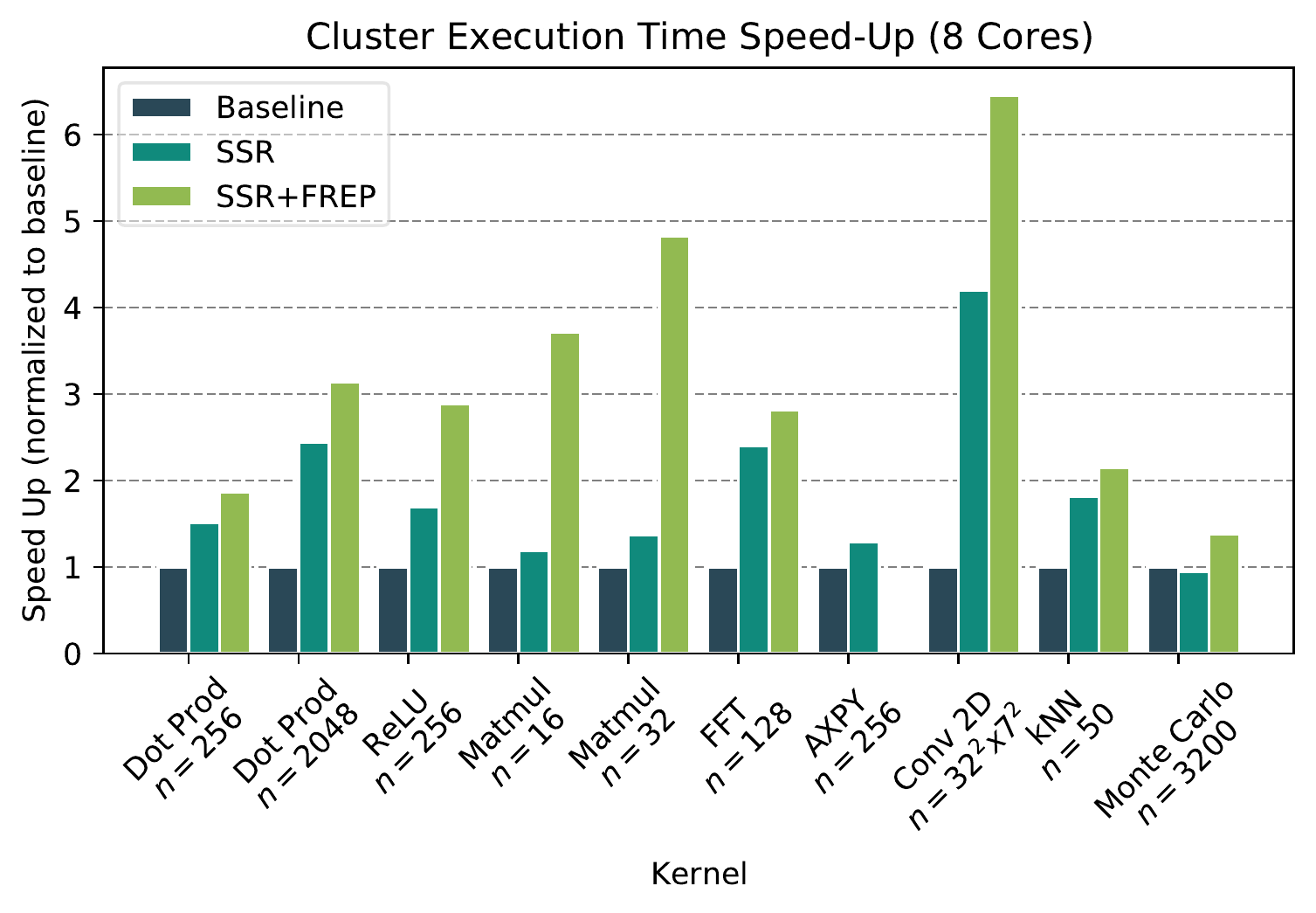}
    \caption{Multi-core speed-up for an octa-core cluster for each microkernel and enabled extension. We can achieve speedups from $1.29 \times$ to $6.45 \times$.}
    \label{fig:multi_core_speed_up}
\end{figure}

\subsubsection{Performance}
\begin{table}
    \caption{FPU Utilization ($\eta$) on a $32 \times 32$ matrix multiplication. Execution time speed-up compared to the single-core baseline ($\Delta$) and speed-up compared to a system with half the cores ($\delta$.)}
    \begin{tabularx}{\linewidth}{@{}rrrrr|rrrrr@{}}
        \toprule
             \# Cores   & $\quad\eta\quad$ & $\quad\delta\quad$ & $\quad\Delta\quad$ & & &\# Cores   &$\quad\eta\quad$   & $\quad\delta\quad$  & $\quad\Delta\quad$  \\\midrule
                     1  &  0.89   & 1.00     & 1.00    & & & 8         &  0.87    & 2.00      & 7.80                  \\
                     2  &  0.90   & 1.98     & 1.98    & & & 16        &  0.81    & 1.87      & 14.62                 \\
                     4  &  0.87   & 1.97     & 3.91    & & & 32        &  0.82    & 1.89      & 27.61                 \\
        \bottomrule
    \end{tabularx}
    \label{tab:par_speedup}
\end{table}

\begin{table}
    \begin{threeparttable}
        \caption{Normalized achieved performance between compute-equivalent Snitch Cluster, Ara \cite{cavalcante2019ara},
        and Hwacha~\cite{dabbelt2016vector} instances for a matrix multiplication, with different
        $n \times n$ problem sizes.}

        \begin{tabular}{@{}rccccccccc@{}}
            \toprule
            $\Pi$ & \multicolumn{3}{c}{4 \glspl{fpu} } & \multicolumn{3}{c}{8 \glspl{fpu} } & \multicolumn{3}{c}{16 \glspl{fpu} } \\
            \cmidrule(lr){1-1} \cmidrule(lr){2-4} \cmidrule(lr){5-7} \cmidrule(lr){8-10}
            $n$   & Snitch & Ara  & Hwacha\tnote{*} & Snitch & Ara   & Hwacha  & Snitch & Ara  & Hwacha \\\midrule
            16    & 68.2   & 49.5 & ---             & 63.2   & 25.4  & ---     & 58.3   & 12.8 & ---    \\
            32    & 87.1   & 82.6 & 49.9            & 84.8   & 53.4  & 35.6    & 81.4   & 27.6 & 22.4   \\
            64    & 93.4   & 89.6 & ---             & 91.7   & 77.5  & ---     & 89.0   & 45.6 & ---    \\
            128   & 96.0   & 94.3 & ---             & 94.7   & 93.1  & ---     & 94.1   & 78.8 & ---    \\\bottomrule
        \end{tabular}
        \begin{tablenotes}
            \item [*] Performance results extracted from~\cite{dabbelt2016vector}
        \end{tablenotes}
    \label{tab:fpu_utilization}
    \end{threeparttable}
\end{table}

For the multi-core performance evaluations we have instantiated an eight core cluster with \SI{8}{\kibi\byte} of instruction cache and \SI{128}{\kibi\byte} of \gls{tcdm} memory (see \figref{fig:die_shot}).
\paragraph{Parallelization}
We have parallelized our kernels to distribute work evenly on all cores. Synchronization between cores is achieved using \riscv's atomic extension and support for atomics on the \gls{tcdm} and on \gls{axi} using AXI5's atomic extension and an atomic adapter~\cite{kurthatuns}. Depending on the workload, parallelization achieves a speed-up from $3\times$ up to $8\times$ for the measured octa-core cluster compared to the single-core version (see \figref{fig:multi_core_vs_single_core_speed_up}). Ideal speed-ups of eight are achieved for the pure \gls{ssr} 2D convolution and the kNN baseline. High multi-core speed-ups can be achieved for matrix multiplication, 2D convolution, kNN, and Monte Carlo methods. The \gls{fft}, dot product and AXPY kernels do not scale that well, mostly due to small problem size which amplifies the reduction and synchronization impact on the overall runtime.
\paragraph{Multi-core speed-up with \gls{ssr} and \gls{frep}}
As can be seen in \figref{fig:multi_core_speed_up} we achieve speed-ups from $1.29 \times$ to $6.45 \times$ depending on the benchmark. As in the single-core case we can use the proposed \gls{ssr} and \gls{frep} extensions to elide explicit load/stores and control flow instructions. In contrast to the single-core case (\figref{fig:single_core_speed_up}) we can observe a slight reduction in speed-up as operand values are potentially (temporarily) unavailable due to contentions on the shared \gls{tcdm} (\gls{sram} bank conflicts), as well as effects of Amdahl's law. Furthermore, we achieve over \SI{94}{\percent} \gls{fpu} utilization for matrices of size $128 \times 128$. As can be seen in \tabref{tab:fpu_utilization} we significantly, by a factor of $4.5$, outperform existing vector processors on small matrix multiplication problems. On larger problems we can show equal or better performance.

The \gls{fft} benchmark demonstrates that the proposed \gls{isa} extensions are also applicable on less linear problems such as \gls{fft}. While we see a decreased \gls{fpu} utilization in the multi-core system (\tabref{tab:par_speedup}) we can observe a total speed-up of $2.8\times$. The decreased \gls{fpu} utilization is attributable to the less linear access pattern and the higher core synchronization frequency for each \gls{fft} stage, which in turn leads to higher contentions as cores are forced to start fetching at the same time from the same memory bank upon each (re-)synchronization.

The Monte Carlo problem is interesting as the pure \gls{ssr} version is slower than the baseline. This is attributed to the fact that the problem needs to re-formulated to operate on blocks of random input data to be beneficial to the streamer infrastructure. The block-wise operation in contrast exhibits \fp data dependencies which could have been filled with integer instructions in the baseline case. Finally, the introduction of \gls{frep} can then fully exploit the fact that integer and \fp pipeline can be executed in parallel exhibiting \emph{pseudo-dual-issue} behavior. The algorithm is still dominated by the integer core generating good random numbers which effectively limits the overall speed-up (we use the xoshiro128+ linear pseudorandom number generator introduced by Blackman and Vigna~\cite{blackman2018scrambled}). The \riscv bit-manipulation extension or a \gls{sfu} dedicated to generating (good) random numbers could significantly enhance this kernel's speed-up.
\subsubsection{Area}
\label{subsec:area}
While the impact of the \gls{frep} extension is confined to \gls{cc} the \gls{ssr} extension also has a cluster-level impact. With \gls{ssr} enabled, each core has two ports into the \gls{tcdm}, increasing the area of the fully connected interconnect. In the selected implementation of an eight-core cluster, we have 16 request ports and 32 memory banks (providing a banking-factor of two). With \SI{155}{\kGE} the \gls{tcdm} interconnect occupies \SI{5}{\percent} of the overall area. The complexity of the crossbar scales with the product of its master and slave ports. We have estimated the complexity of a \num{32} requests and \num{64} banks crossbar to be around \SI{630}{\kGE} and the area of a \num{64} request ports and \num{128} banks to be around \SI{2.5}{\mega\gateequivalent}.

\subsubsection{Energy Efficiency and Power}
\begin{figure}
    \centering
    \includegraphics[width=\linewidth]{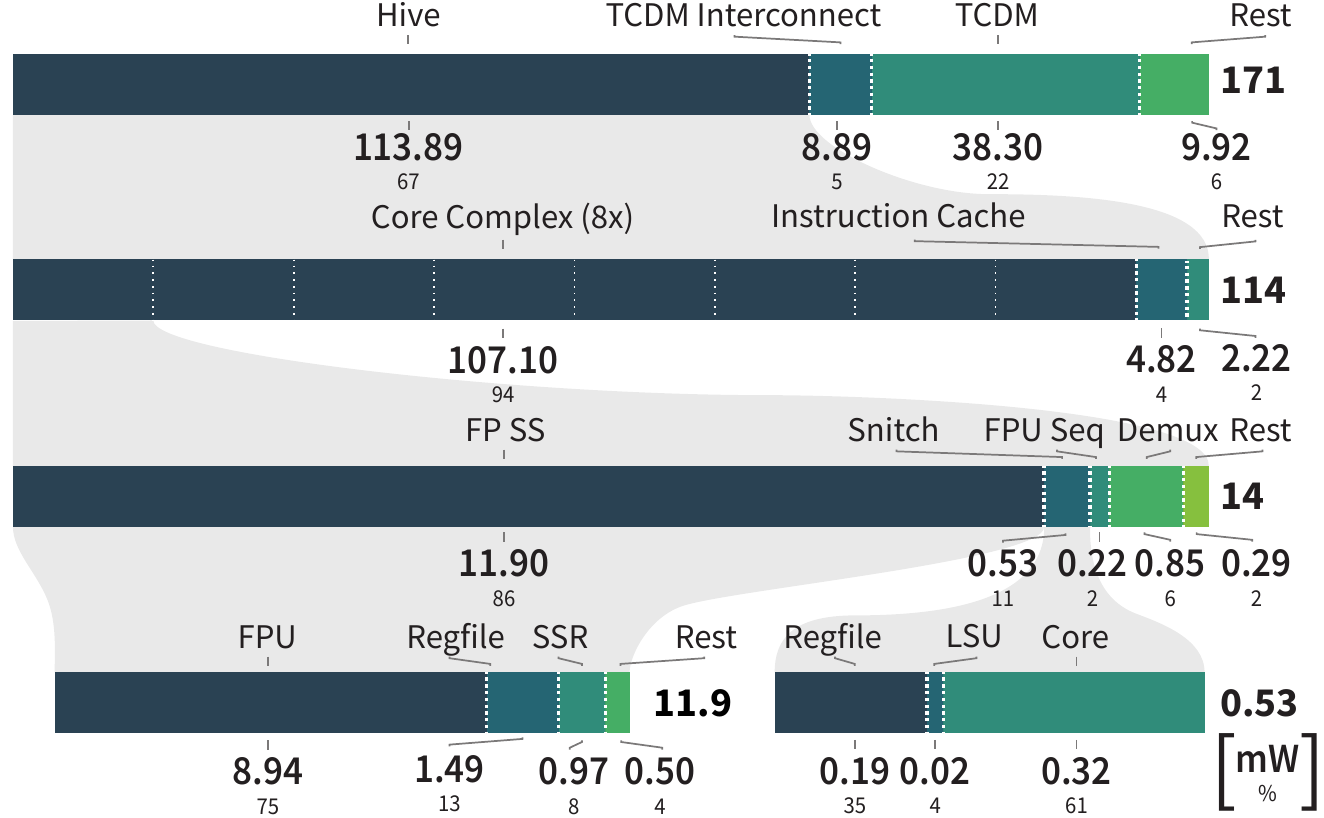}
    \caption{Hierarchical power distribution estimates obtained using \primetime at \SI{1}{\giga\hertz} and \SI{25}{\degreeCelsius} on a 32$\times$32 matrix multiplication kernel using the proposed \gls{ssr} and \gls{frep} extensions. All integer core only use \SI{1}{\percent} of the overall power. The necessary hardware for the \glspl{ssr} and the \gls{frep} extension uses less than \SI{4}{\percent} and \SI{1}{\percent} of the total power respectively. The Snitch core has been configured with RV32I with an FF-based \gls{rf} and \glspl{pmc}}
    \label{fig:power_bowtruckle}
\end{figure}
\begin{figure}
    \centering
    \includegraphics[width=\linewidth]{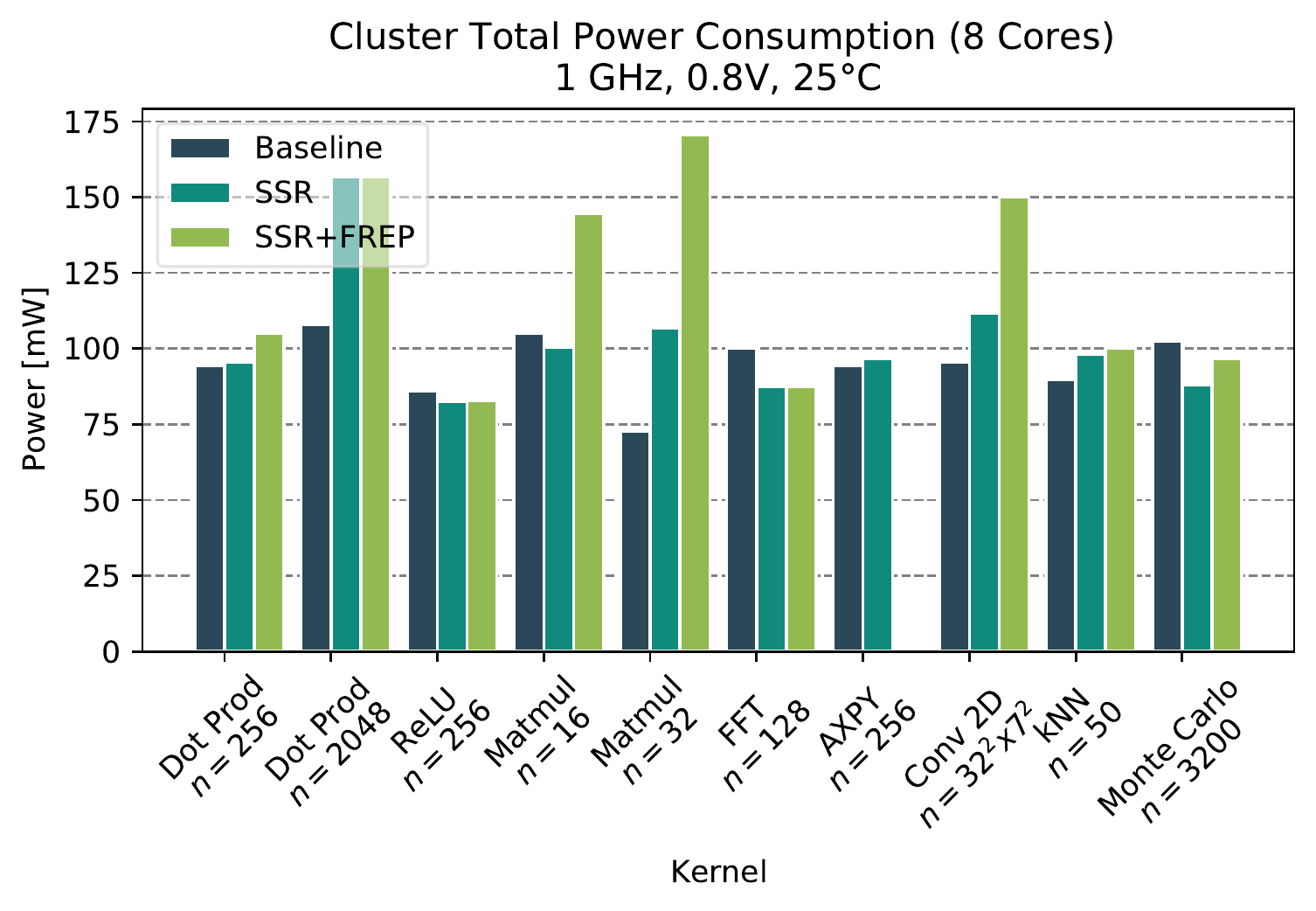}
    \caption{Power consumption of an octa-core cluster for all microkernels and proposed \gls{isa} extensions.}
    \label{fig:multi_core_power}
\end{figure}
\begin{figure}
    \centering
    \includegraphics[width=\linewidth]{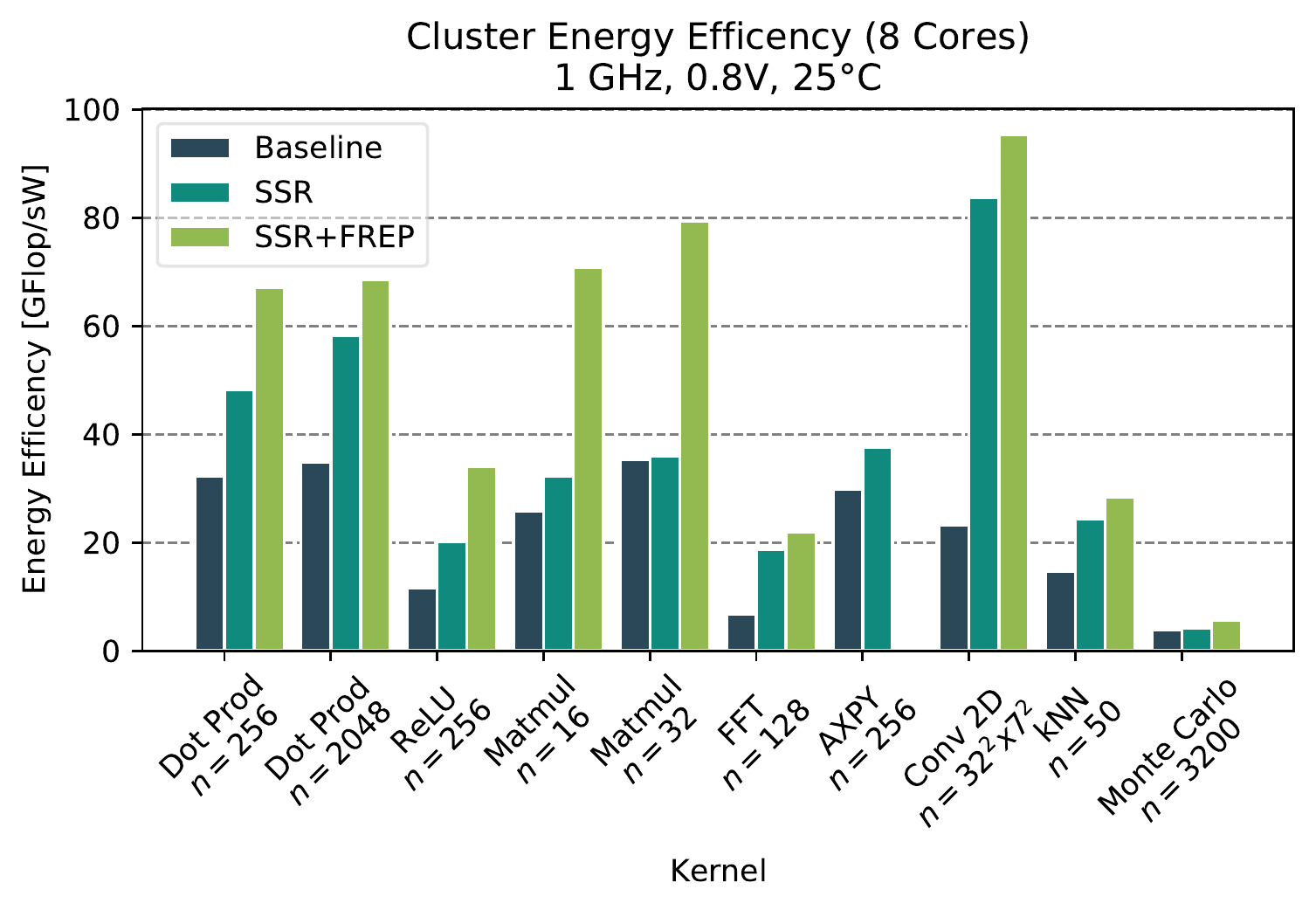}
    \caption{Energy efficiency of an octa-core cluster for all microkernels and proposed \gls{isa} extensions. The proposed cluster architecture achieves up to \SI{80}{\GFLOPsW} peak energy efficiency at \SI{1}{\giga\hertz}, \SI{0.8}{\volt} and \SI{25}{\degreeCelsius}. For the different kernels we achieve an increase of \numrange{1.5}{4.9} in energy efficiency. The Monte Carlo benchmark offers a poor energy-efficiency per flop as the generation of good random numbers takes up significant amounts of energy (we use the xoshiro128+ algorithm for fast \fp number generation \cite{blackman2018scrambled}).}
    \label{fig:multi_core_eficiency}
\end{figure}

We have selected a $32 \times 32$ matrix multiplication benchmark running on a post-layout netlist to give an indicative power break-down of the system's component (\figref{fig:power_bowtruckle}). For the given benchmark the cluster consumes a total of \SI{171}{\milli\watt} of which \SI{63}{\percent} are consumed in the \gls{cc}, \SI{5}{\percent} in the interconnect and \SI{22}{\percent} in the \gls{sram} banks of the \gls{tcdm}. \SI{42}{\percent} of the energy is spent in the actual \gls{fpu} on the computation. While the integer control core only uses \SI{1}{\percent} of the overall power. The additional hardware for \gls{ssr} and \gls{frep} only make up for a fraction of the overall power consumption, less than \SI{4}{\percent} and \SI{1}{\percent} respectively. What is particularly interesting ist that the instruction cache only consumes \SI{4.8}{\milli\watt} or \SI{4}{\percent} of the total cluster power. This is due to the \gls{frep} extension servicing the \gls{fpu} from its local sequence buffer, and the Snitch integer core exhibiting a very low activity that can mostly be served from its L0 instruction cache, that has been implemented as a FF-based memory and can be read and written using less energy compared to \glspl{sram}. The total power of all micro-benchmarks is given in \figref{fig:multi_core_power}. As we only see a marginal increase in power for the given benchmarks but a significant improvement in execution speed and a high \gls{fpu} utilization we can observe a similiar increase in energy efficiency. \figref{fig:multi_core_eficiency} shows a \numrange{1.5}{4.1} increase in energy efficiency compared to the baseline. The systems achieves an absolute peak energy efficiency of close to \SI{80}{\DGFLOPsW} and \SI{104}{\SGFLOPsW} for double precision matrix multiplication and up to \SI{95}{\DGFLOPsW} for the 2D convolution benchmark.

To put the absolute energy efficiency into perspective, we estimated the achievable peak energy efficiency in \SI{22}{\nano\meter}. Every architecture, even highly specialized accelerators, must at least perform two loads and a \gls{fma} instruction for each element. We can, therefore, estimate the energy-efficiency upper bound of \SI{120}{\DGFLOPsW}.  Snitch achieves \SI{79}{\percent} of this theoretical peak efficiency.

\section{Related Work}
\begin{table}
    \begin{threeparttable}
    \caption{Comparison with Ara~\cite{cavalcante2019ara} and NVIDIA Xavier SoC~\cite{ditty2018nvidia} on an $n \times n$ matrix multiplication.}
    \begin{tabularx}{\linewidth}{@{}lXrrr|rrr@{}}
        \toprule
                                          &                                       & Snitch  & Ara                      &&  Volta SM              & \,Carmel\tnote{*}       \\
                                          & Unit                                  & Us      & \cite{cavalcante2019ara} && \cite{ditty2018nvidia} & \cite{ditty2018nvidia}  \\ \midrule
            Problem Size                  & $n$                                   & 32      & 32                       && 256                    &  256                    \\
            Base \gls{isa}                &                                       & RV      & RV                       && Volta                  & \arm                    \\
            Technode                      & [\si{\nano\meter}]                    & 22      & 22                       && 12                     &  12                     \\
            Clock (typical)               & [\si{\giga\hertz}]                    & 1.06    & 1.17                     && 1.38                   &  2.27                   \\
            Clock (worst)                 & [\si{\giga\hertz}]                    & 0.75    & 0.87                     && ---                    &  ---                    \\
            Peak SP                       & [\si{\GFLOPs}]                        & 16.96   & 18.72                    && 176                    & 36.25                   \\
            Peak DP                       & [\si{\GFLOPs}]                        & 16.96   & 18.72                    && \tnote{$\dagger$}---   & 18.13                   \\
            Sustained SP                  & [\si{\GFLOPs}]                        & 14.38   & 10.00                    && \tnote{$\ddagger$}153  & \tnote{\S}22.10         \\
            Sustained DP                  & [\si{\GFLOPs}]                        & 14.38   & 10.00                    && \tnote{$\dagger$}---   & \tnote{$\parallel$}9.27 \\
            Utilization SP                & [\si{\percent}]                       & 84.80   & ---                      && 86.66                  & 60.97                   \\
            Utilization DP                & [\si{\percent}]                       & 84.80   & 53.40                    && \tnote{$\dagger$}---   & 51.15                   \\
            Impl. Area\tnote{\#}          & [\si{\square\milli\meter}]            & 0.89    & 1.07                     && 11.03                  & \tnote{**}7.37          \\
            Area Eff. SP                  & [\si{\GFLOPs\per\square\milli\meter}] & 25.83	& ---                      && 13.84                  & 3.00	                   \\
            Area Eff. DP                  & [\si{\GFLOPs\per\square\milli\meter}] & 25.83	& 17.53                    && 13.84                  & 1.26                    \\
            Tot. Power SP                 & [\si{\watt}]                          & 0.13    & ---                      && 2.91                   & 2.16                    \\
            Tot. Power DP                 & [\si{\watt}]                          & 0.17    & 0.46                     && \tnote{$\dagger$}---   & 1.85                    \\
            Leakage                       & [\si{\milli\watt}]                    & 12      & 21.1                     && ---                    & ---                     \\
            Energy Eff. SP                & [\si{\GFLOPsW}]                       & 103.84  & ---                      && 52.39                  & 10.24                   \\
            Energy Eff. DP                & [\si{\GFLOPsW}]                       & 79.42   & 39.9                     && \tnote{$\dagger$}---   & 5.01                    \\
        \bottomrule
    \end{tabularx}
    \begin{tablenotes}
        \item [*] Single-core, estimated from the eight core core complex including L3 cache \\
        \item [$\dagger$] The Volta SM in Tegra Xavier does not contain any double precision \glspl{fpu} \\
        \item [$\ddagger$] Measured using the SGEMM implementation of CUBLAS~\cite{nvidia2007cublas} \\
        \item [\S] Measured using an SGEMM implementation of the ARM ComputeLibary using NEON \gls{isa} extension~\cite{compute2019arm} \\
        \item [$\parallel$] Measured using the OpenBLAS implementation~\cite{xianyi2012openblas} \\
        \item [\#] Post-layout area measured from die photograph \\
        \item [**] Including proportionate L2 and L3 caches  \\
    \end{tablenotes}
    \label{tab:rel_work}
    \end{threeparttable}
\end{table}

The problem of keeping the \gls{fpu} utilization high has been the subject of a lot of architecture research. The most prominent and widely used techniques encompass super-scalar (out-of-order), general-purpose, \glspl{cpu}, (Cray-style) vector architectures and general-purpose compute using \glspl{gpu}. While these architectures promise to deliver high performance, they do not target energy efficiency as their primary design goal.

\subsection{Vector Architectures}
\label{subsec:vector_arch}
Cray-style vector architectures are enjoying renewed popularity with \arm providing their \gls{sve}~\cite{stephens2017arm} and \riscv actively developing a vector extension~\cite{vector2020riscv}. An early, but complete version of the \riscv vector extension in \SI{22}{\nano\meter} called Ara, has been implemented by Cavalcante \emph{et~al.}~\cite{cavalcante2019ara}. The same technology node and configuration size allow for a direct comparison to our architecture. As a comparison point, we chose an eight-lane configuration that delivers a peak of \SI{16}{\DFLOP\per\cycle} equal to the octa-core cluster we have presented in the evaluation section. The vector architecture accelerates programs that work on vectored data by providing a single-instruction which operates on (parts of) the vector. The instruction front-end of the attached core is feeding the vector unit special vector instructions that can then independently operate on chunks of data from the vector register file. The vector register file is similar in size and access latency to the \gls{tcdm} in a Snitch cluster. However, in stark contrast to the vector register file, our system allows us to access individual elements of the \gls{tcdm} as it is byte-wise addressable. The vector architecture compensates this fact by providing dedicated shuffle instructions, which, in contrast, consume precious instruction bandwidth and issue-slots.

As a consequence, the scalar core needs to issue many instructions to the vector architecture that potentially bottleneck the instruction front-end and hence performs poorly on smaller and finer granular problems (see \tabref{tab:fpu_utilization}). On smaller matrix multiplication problems, our architecture significantly outperforms, by a factor of 4.5, the Ara vector architecture as our \gls{tcdm} interconnect and byte-wise access to the \gls{tcdm} provides implicit shuffle semantic. On increasing problem sizes, the vector architecture catches up in performance, but we can retain superiority even for larger problem sizes (see \tabref{tab:fpu_utilization}).

The rigid, linear access pattern, superimposed by the nature of vectors, imposes yet another problem: To compensate for the lack of access semantic into the register file additional \gls{isa} extensions such as 2D and tensor extensions are needed to encode the more complicated access patterns. As the shape of the computation is encoded in the instruction, this significantly bloats the encoding space, which in turn makes the instruction-frontend and decoding logic more complex and hence more energy-inefficient. In contrast the \gls{ssr} and \gls{frep} extension provide up to 4 access dimensions in their current implementation. With the implicit load/store encoding into register reads/writes, no new instructions are needed, and the instruction-frontend and decoding logic is identical to the scalar core.

\tabref{tab:rel_work} compares several figures of merit between Ara (Ariane's vector extension) and the same size Snitch system. Both systems offer the same number of floating-point operations per cycle at comparable clock-frequency. On the chosen problem size of a $32 \times 32$ matrix multiplication, our system offers more than $1.5 \times$ sustained \fp operations at twice the energy efficiency of almost \SI{80}{\GFLOPsW} compared to \SI{40}{\GFLOPsW} of Ara. A similar comparison can be done for the axpy and 2D convolution benchmark, where we achieve $2.45\times$ and $2.37\times$ the energy efficiency improvement over Ara. Most of the energy efficiency gains come from the higher area efficiency and the much higher compute/control ratio. A comparable architecture to Ara is Hwacha~\cite{dabbelt2016vector}, which suffers from similar limitations.

\subsection{\Glspl{gpu}}
\Glspl{gpu} have completely penetrated the market of general-purpose computing with their superior capabilities to accelerate dense linear algebra kernels most prominently found in machine-learning applications. The key idea of \gls{gpgpu} is to oversubscribe the compute units using multiple, parallel threads that can be dynamically scheduled by hardware to hide access latencies to memory. We have estimated energy efficiency of an NVIDIA \gls{gpu} using a Tegra Xavier \gls{soc}~\cite{ditty2018nvidia} development kit. The board allows for direct power measurements on the supply rails of both the \gls{gpu} and \gls{cpu}. The Tegra \gls{soc} contains a Volta-based~\cite{nvidia2017nvidia} \gls{gpu} consisting of eight \glspl{sm} which each in turn consists of 32 double- and 64 single-precision \glspl{fpu}. Each \gls{sm} contains four execution units, each managing eight double-precision and 16 single-precision \glspl{fpu}, which share a common register file and an instruction cache. Hence such a quadrant is directly comparable to one Snitch cluster as presented here. Clock speeds of \SI{1}{\giga\hertz} of Snitch and \SI{1.38}{\giga\hertz} for the Volta \gls{sm} are comparable keeping in mind that the \gls{sm} has been manufactured in a more advanced technology, see~\tabref{tab:rel_work}. On a high-level comparison, the Snitch system surpasses the \gls{sm} in terms of energy efficiency, by over $1.98$ on single-precision workloads. This comparison does not take technology scaling into consideration, which would further improve energy-efficiency in favor of Snitch.

\subsection{Super-scalar \glspl{cpu}}
The Tegra Xavier \gls{soc} also offers an eight-core cluster of NVIDIA's ARMv8 implementation called Carmel. The Carmel \gls{cpu} is a 10-issue, super-scalar \gls{cpu} including support for \arm's \gls{simd} extension NEON. Each core contains two 128-bit \gls{simd}-\glspl{fpu} that are fracturable in either two 64-bit, four 32-bit or eight 16-bit units, offering a total of 8 double-precision \si{\FLOP\per\cycle}, hence comparable to the presented octa-core Snitch cluster. The processor runs at a substantially higher clock frequency of \SI{2.27}{\giga\hertz} at the expense of a much deeper pipeline, which in turn requires the processor to hide pipeline stalls by exploiting \gls{ilp} in the form of super-scalar execution and a steep memory hierarchy to mitigate the effects of high memory latency. The increased hardware cost reduces the attainable area efficiency to only \SI{1.26}{\DGFLOPs\per\square\milli\meter}. The losses in area efficiency have a direct influence on the energy efficiency of the system. Not accounting for technology scaling, we can show more than $10\times$ improvement in energy efficiency for FP32 and $15\times$ for FP64.

Recent developments in high-performance chips, such as Fujitsu's A64FX~\cite{yoshida2018fujitsu}, clearly demonstrate that energy-efficiency is becoming the number one design concern. The new Green500~\cite{feng2007green500} winner achieves \SI{16.876}{\DGFLOPsW} system-level energy-efficiency (including cooling, board and power supplies). Unfortunately, as we do not have access to such a system for detailed measurements, we can not perform accurate direct comparisons.

\section{Conclusion}
We present a general-purpose computing system tuned for the highest possible energy efficiency on double-precision \fp arithmetic. The system offers an implementation of the \riscv atomic extension (A) for efficient multi-core programming and can be targeted with a standard \riscv toolchain. We outperform existing state-of-the-art systems (\tabref{tab:rel_work} on energy efficiency by a factor of $2$ by leveraging several ideas.
\paragraph*{Tightly Coupled Data Memory (TCDM)}
Explicit scratchpad memories (\gls{tcdm}) instead of hardware managed caches enable deterministic data placement and avoid suboptimal cache replacement strategies. The \gls{tcdm} memory is shared amongst a cluster of cores, making data sharing significantly more energy efficient as no cache coherence protocol is necessary.
\paragraph*{Small and efficient integer core}
We aim to maximize the control to compute ratio by providing a small and agile integer core that can do single-cycle control flow decisions and integer arithmetic and combine it with a large \gls{fpu}. The \gls{fpss} decouples the integer/control flow from the \fp operations and the \gls{fpss} can operate on its own register file and provides its own \gls{fp} \gls{lsu}.
\paragraph*{\gls{isa} extensions}
We provide two minimal impact \gls{isa} extensions, \glspl{ssr} and \gls{frep}. The first makes it possible to set up a four-dimensional stream to memory from which the core can simply read/write using two designated register names. The \gls{frep} extension complements the \gls{ssr} extension by further decoupling the issuing of \fp instructions to the \gls{fpss}. The integer core pushes \riscv instructions into the previously configured loop-buffer and subsequently issue those instructions to the \gls{fpu}. This has two beneficial side-effects: While the \gls{fpu} loop-buffer feeds the \gls{fpu} with instructions, the integer core is free to do auxiliary tasks, such as orchestrating data movement. The second positive effect is that it relieves the pressure on the instruction cache, therefore saving energy.

The system achieves a speed-up of up to $6.45\times$ on data-oblivious kernels while still being fully programmable and not overspecializing on one problem domain. The flexibility offered by the small, integer control unit makes it a versatile architecture and possible to adapt to changing algorithmic requirements. Furthermore, we have shown that eight cores per cluster provide a good trade-off between speed-up and complexity of the interconnect (see \tabref{tab:par_speedup} and \secref{subsec:area}). A future extension of the proposed \gls{ssr} hardware could target improved efficiency for sparse linear algebra problems. Furthermore, extended benchmarking and improvements in the compiler infrastructure are exciting future research directions.


%



\ifCLASSOPTIONcompsoc
  \section*{Acknowledgments}
\else
  \section*{Acknowledgment}
\fi

This work has received funding from the European Union’s Horizon 2020 research and innovation programme under grant agreement number 732631, project ``OPRECOMP''.

\ifCLASSOPTIONcaptionsoff
  \newpage
\fi



\bibliographystyle{IEEEtran}
\bibliography{IEEEabrv,refs}

\vskip -2\baselineskip plus -1fil

\begin{IEEEbiography}[{\includegraphics[width=1in,height=1.25in,clip,keepaspectratio]{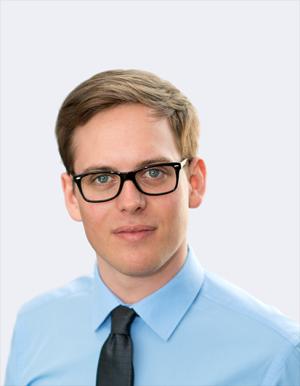}}]{Florian Zaruba}
  received his BSc degree from TU Wien in 2014 and his MSc from the Swiss Federal Institute of Technology Zurich in 2017.  He is currently pursuing a PhD degree at the Integrated Systems Laboratory.  His research interests include design of very large scale integration circuits and high performance computer architectures.
\end{IEEEbiography}

\vskip -2\baselineskip plus -1fil

\begin{IEEEbiography}[{\includegraphics[width=1in,height=1.25in,clip,keepaspectratio]{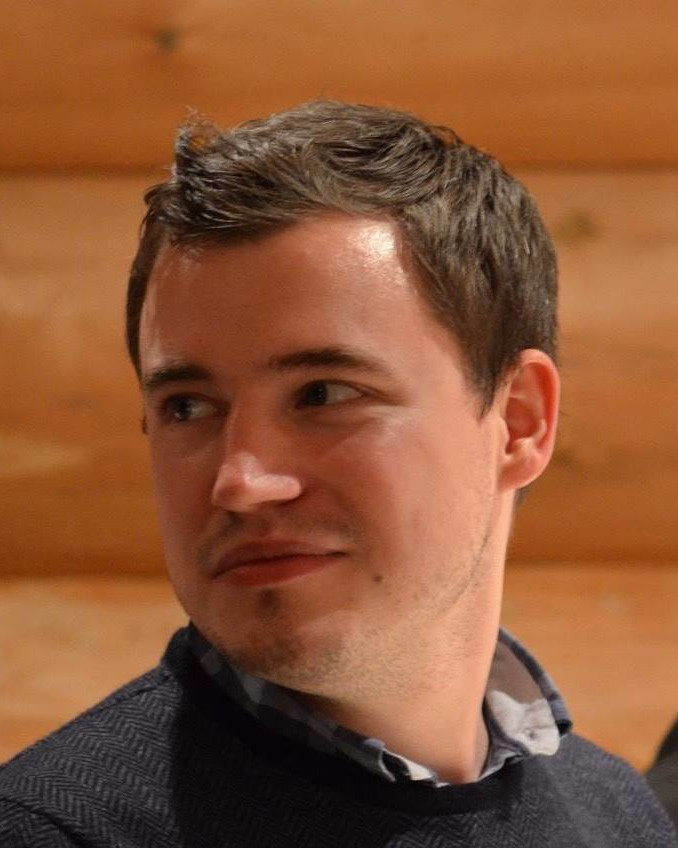}}]{Fabian Schuiki}
  received the B.Sc. and M.Sc. degree in electrical engineering from ETH Zürich, in 2014 and 2016, respectively.  He is currently pursuing a Ph.D. degree with the Digital Circuits and Systems group of Luca Benini.  His research interests include computer architecture, transprecision computing, as well as near- and in-memory processing.
\end{IEEEbiography}

\vskip -2\baselineskip plus -1fil

\begin{IEEEbiography}[{\includegraphics[width=1in,height=1.25in,clip,keepaspectratio]{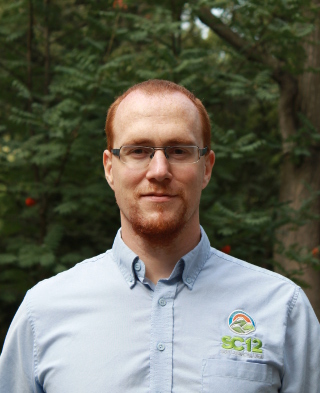}}]{Torsten Hoefler}
  is a Professor of Computer Science at ETH Zürich, Switzerland. He is also a key member of the Message Passing Interface (MPI) Forum where he chairs the ``Collective Operations and Topologies'' working group.  His research interests revolve around the central topic of ``Performance-centric System Design'' and include scalable networks, parallel programming techniques, and performance modeling.  Torsten won best paper awards at the ACM/IEEE Supercomputing Conference SC10, SC13, SC14, EuroMPI'13, HPDC'15, HPDC'16, IPDPS'15, and other conferences.  He published numerous peer-reviewed scientific conference and journal articles and authored chapters of the MPI-2.2 and MPI-3.0 standards.  He received the Latsis prize of ETH Zurich as well as an ERC starting grant in 2015.
\end{IEEEbiography}

\vskip -2\baselineskip plus -1fil

\begin{IEEEbiography}[{\includegraphics[width=1in,height=1.25in,clip,keepaspectratio]{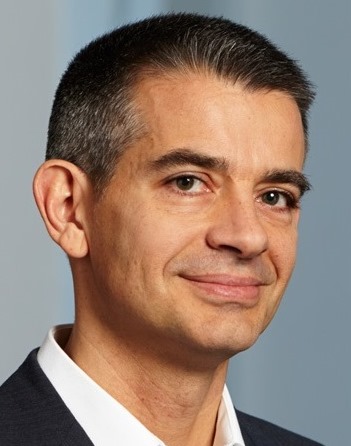}}]{Luca Benini}
  holds the chair of digital Circuits and systems at ETHZ and is Full Professor at the Universita di Bologna.  Dr. Benini's research interests are in energy-efficient computing systems design, from embedded to high-performance.  He has published more than 1000 peer-reviewed papers and five books.  He is a Fellow of the ACM and a member of the Academia Europaea. He is the recipient of the 2016 IEEE CAS Mac Van Valkenburg award.
\end{IEEEbiography}

\end{document}